\documentclass[12pt,twoside]{article}

\pdfoutput=1

\usepackage[usenames]{color}
\usepackage{lscape}
\usepackage{amssymb}
\usepackage{rotating}
\usepackage[T1]{fontenc}
\usepackage{graphicx}
\usepackage{fancyhdr}
\usepackage{amsmath}
\usepackage{ae}
\usepackage{aecompl}
\usepackage{hyperref}
\usepackage{subfigure}

\def\nn{\nonumber \\}
\def\ov{\overline}

\def\ptmiss{\ensuremath{\vec{p}_\mathrm{T}^{\rm miss}}}
\def\Etmiss{\ensuremath{E_\mathrm{T}^{\rm miss}}}
\def\ptvec{\ensuremath{\vec{p}_\mathrm{T}}}
\def\pt{\ensuremath{p_\mathrm{T}}}
\def\meff{\ensuremath{m_{\rm eff}}}

\newlength{\dinwidth}
\newlength{\dinmargin}
\begin{document}

\thispagestyle{empty}

\begin{flushright}
Cavendish--HEP--11/25\\
DAMTP--2011--110\\
DESY 11-256
\end{flushright}

\vspace*{1cm}

\centerline{\Large\bf LHC constraints on Yukawa unification in SO(10)}

\vspace*{5mm} \noindent
\vskip 0.5cm
\centerline{\bf
Marcin Badziak${}^{a,b,}$\footnote[1]{M.Badziak@damtp.cam.ac.uk},
Kazuki Sakurai${}^{c}$\footnote[2]{kazuki.sakurai@desy.de}
}
\vskip 5mm
\begin{center}
{${}^a$\em Department of Applied Mathematics and Theoretical Physics, 
Centre for Mathematical Sciences, University of Cambridge, 
Wilberforce Road, Cambridge CB3 0WA, United Kingdom \\[5mm]

${}^b$Cavendish Laboratory, University of Cambridge, J.J. Thomson Avenue, \\ Cambridge CB3 0HE, United Kingdom \\[5mm]

${}^c$ Deutsches Elektronen Synchrotron DESY, Notkestrasse 85, D-22607
Hamburg, Germany}
\end{center}

\vskip 1cm

\centerline{\bf Abstract}
\vskip 3mm

LHC constraints on the recently proposed SUSY SO(10) GUT model with top-bottom-tau Yukawa unification 
are investigated.
In this model, various phenomenological constraints are in concord with the Yukawa unification 
thanks to the negative sign of $\mu$, $D$-term splitting in the soft scalar masses and non-universal gaugino masses
generated by a non-zero $F$-term in a 24-dimensional representation
of SU(5) $\subset$ SO(10).
After discussing the impact of the CP-odd Higgs boson mass bound on this model,
we provide a detailed  analysis of the recent direct SUSY searches performed by ATLAS
and investigate the constraints on this SO(10) model.
At 95\% confidence level, the lower limit on the gluino mass is found to be 675 GeV.
Assuming an integrated luminosity of 10 fb$^{-1}$, this bound may be extended to 1.1 TeV if the right-handed down squark is lighter than about 1 TeV.

\newpage

%%%%%%%%%%%%%%%%%%%%%%%%%%%%%%%%%%%%%%%%%%%%%%%%%%%%%%%%%%%%%%%%%%%%%%
%%%%%%%%%%%%%%%%%%%%%%%%%%%%%%%%%%%%%%%%%%%%%%%%%%%%%%%%%%%%%%%%%%%%%%

%%%%%%%%%%%%%%%%%%%%%%%%%%%%%%%%%%%%%%%%%%%%%%%%%%%%%%%%%%%%%%%%%%%%%%
%%%%%%%%%%%%%%%%%%%%%%%%%%%%%%%%%%%%%%%%%%%%%%%%%%%%%%%%%%%%%%%%%%%%%%

\section{Introduction}

In the simplest versions of supersymmetric (SUSY) Grand Unified Theories (GUTs) based on the SO(10) gauge symmetry, in which the Minimal Supersymmetric Standard Model (MSSM) Higgs doublets, $H_u$ and $H_d$, reside in a $\bf{10}$ of SO(10), the Yukawa couplings of top, bottom and tau unify at the GUT scale \cite{tbtau}. Unlike the gauge coupling unification, Yukawa coupling unification is very sensitive to weak scale threshold corrections which  depend on the entire SUSY spectrum \cite{Hall,Carena,Pierce}. Moreover, top-bottom-tau Yukawa unification is often in conflict with radiative electroweak symmetry breaking (REWSB)  \cite{Carena}. 
Therefore, the condition of the top-bottom-tau Yukawa unification at the GUT scale imposes non-trivial constraints on the soft SUSY breaking terms. 
In particular, it has been shown that the Yukawa unification is incompatible with the REWSB in the Constrained MSSM (CMSSM) where the soft masses for scalars and gauginos are universal at the GUT scale \cite{Carena}.

SO(10) models are rather constrained not only because of the requirement of Yukawa unification but also because all the MSSM sfermions of each generation fit into a single $\bf{16}$ of SO(10). Thus, 
assuming the Minimal Flavor Violation \cite{MFV1, MFV2}, 
the only non-universality in the scalar masses consistent with the SO(10) gauge symmetry is the mass splitting between sfermions and Higgses. This has been shown to be insufficient to make the Yukawa unification compatible with the REWSB \cite{Olechowski}. However, if the SO(10) gauge symmetry is spontaneously broken, as is generally expected, in the effective theory below the GUT scale, a generic $D$-term contribution splits the scalar masses residing in the same representation of SO(10) \cite{Dterm}. 
Most importantly, the $D$-term contribution splits the masses of the MSSM Higgs doublets and may 
make the Yukawa unification consistent with the REWSB \cite{Murayama}.

The requirement of the top-bottom-tau Yukawa unification prefers negative values of the Higgs-mixing parameter, $\mu$.\footnote{We use the sign convention in which the gluino mass parameter $M_3$ and the gluino mass $m_{\tilde{g}}$ are positive.}
However, a model with the negative $\mu$ and the universal gaugino masses \cite{Baernegmu} predicts a 
negative SUSY contribution to the muon anomalous magnetic moment, $(g-2)_{\mu}$, and enlarges the observed discrepancy between theory and experiment.
That is why most of the studies that assume the universal gaugino masses 
in the last decade have been devoted to the case of the positive $\mu$ \cite{HS,DR3}. 
In such a case the Yukawa unification can be realized but there is a price for it. 
Namely, the scalar masses are pushed to the multi-TeV regime so proper REWSB is obtained at the expense of severe fine-tuning. 
Moreover, these models have serious problems with the FCNC constraints \cite{FCNCYukawa} because of a very large value of $|\mu A_t|$, in particular with a lower limit on BR$(B_s \to \mu^+\mu^-)$ which was recently improved by the CMS and the LHCb experiments \cite{BsmumuCMSLHCb}. 

Last but not least, in that class of models the solutions with the most accurate unification of the Yukawa couplings at the GUT scale found in the numerical scans performed in Ref.\,\cite{HS,DR3,testingYUposmu} are characterised by the gluino mass of about 350 GeV \footnote{
The usually dominant SUSY correction to the bottom mass is proportional to $\mu m_{\tilde g}/m_{\tilde b_2}^2$ when $m_{\tilde b_2}  \gtrsim  m_{\tilde g}$. Assuming the Yukawa unification and $\mu>0$, this correction drives the bottom mass away from its experimental value so gluino is required to be much lighter than the heavier sbottom to suppress this correction. Typically this implies very light gluino in accord with findings of Ref.\,\cite{HS,DR3,testingYUposmu}. However, the necessity of very light gluino in the Yukawa-unified models with $\mu>0$ was questioned recently in Ref.\,\cite{YUposmu_heaviergluino} where new solutions with exact Yukawa unification were found with the gluino mass as large as 1.4 TeV.
}, while the recent ATLAS ``b-jet'' search \cite{ATLASbjets} excluded the gluino masses up to 450 (570) GeV in a model with ``ad hoc'' splitting of the Higgs masses \cite{HS} ($D$-term splitting of the scalar masses \cite{DR3}).  For a recent review of SO(10) models with the third family Yukawa coupling unification see Ref.\,\cite{YUreview}.

A remedy to the problems of the SO(10) models discussed above was put forward recently in 
Ref.\,\cite{bop} 
where a novel model with negative $\mu$ was proposed. In this model the gaugino masses are assumed to be generated by an $F$-term which is a non-singlet of SO(10) transforming as a $\bf{24}$ of SU(5) $\subset$ SO(10). Such an assumption leads to the gaugino masses with the following pattern at the GUT scale: $M_1:M_2:M_3=-1:-3:2$ \cite{Martin}. The main phenomenological success of the model is its consistency with all experimental constraints including the $(g-2)_{\mu}$ and the BR$(b \to s \gamma)$ which are satisfied thanks to the positive sign of the product $\mu M_2$ \cite{ChNath}. Another advantage of the model is the prediction of a rather light SUSY spectrum which can be tested at the LHC. In particular, the gluino mass is predicted to be in the range of about $500-700$ or $900-1550$ GeV. It is not unreasonable to expect that some of the parameter space corresponding to such a light gluino has already been excluded by the LHC. 
However, the LHC experiments interpret their data mainly in the CMSSM and some simplified models and the obtained mass limits for SUSY particles cannot be applied directly to other models. 
There are several studies that reinterpret the LHC SUSY searches in terms of the SUSY models 
other than the CMSSM.  
The LHC constraints on gauge mediation \cite{GMSB_LHC1, GMSB_LHC2}, 
anomaly mediation \cite{mAMSB_LHC}, 
phenomenological SUSY models \cite{pheno_LHC, pheno_LHC2}, 
SUSY models with compressed spectra \cite{compressed_LHC} 
and light third generation models 
\cite{thirdgen_LHC1, thirdgen_LHC_add, thirdgen_LHC2, thirdgen_LHC3, thirdgen_LHC4, thirdgen_LHC5} 
have been studied.
The constraints on models with the top-bottom-tau Yukawa unification in the context of SUSY ${\rm SU}(4)_c\times {\rm SU}(2)_L\times {\rm SU}(2)_R$ \cite{GoKhRaSh} have also been investigated \cite{LHC_gluinoNLSP}. 
In the present paper we study the constraints and implications of the recent LHC data on the SO(10) model proposed in Ref.\,\cite{bop}. We focus on three ATLAS analyses of the final states with large missing transverse momentum and no leptons: the ``0-lepton'' search with 1.04 fb$^{-1}$ \cite{ATLAS0lepton}, the ``b-jet'' search with 0.83 fb$^{-1}$ \cite{ATLASbjets} and the ``multijets'' search with 1.34 fb$^{-1}$ of data \cite{ATLASmultijets}. 

The paper is organized as follows. In section \ref{sec_YukF24} we describe in some more detail the SO(10) model proposed in Ref.\,\cite{bop}. In section \ref{sec_ATLAS} we summarise the ATLAS searches for squarks and gluinos. In section \ref{sec_method} we present our methodology of simulating the ATLAS searches and its validation. Interpretation of the ATLAS searches in the SO(10) model proposed in Ref.\,\cite{bop} is given in section \ref{sec_ATLASso10} . We conclude in section \ref{sec_concl}. In the Appendix we give additional information on the validation of our simulation of the ATLAS searches.

%%%%%%%%%%%%%%%%%%%%%%%%%%%%%%%%%%%%%%%%%%%%%%%%%%%%%%%%%%%%%%%%%%%%%%
%%%%%%%%%%%%%%%%%%%%%%%%%%%%%%%%%%%%%%%%%%%%%%%%%%%%%%%%%%%%%%%%%%%%%%

%%%%%%%%%%%%%%%%%%%%%%%%%%%%%%%%%%%%%%%%%%%%%%%%%%%%%%%%%%%%%%%%%%%%%%
%%%%%%%%%%%%%%%%%%%%%%%%%%%%%%%%%%%%%%%%%%%%%%%%%%%%%%%%%%%%%%%%%%%%%%

\section{Description of the SO(10) model}
\label{sec_YukF24}

We start with a brief review of the SO(10) model proposed in Ref.\,\cite{bop}. The first crucial feature of this model is that the Higgs-mixing parameter $\mu$ is negative. 
The negative sign of $\mu$ is phenomenologically preferred because it allows for the negative threshold correction to the bottom mass \cite{Hall,Carena,Pierce} (as required by the Yukawa unification) without the need to push up the scalar masses to multi-TeV regime. Even though negative $\mu$ is preferred by the Yukawa unification, it leads to discrepancy with the experimental results for the $(g-2)_{\mu}$ and the BR$(b\to s \gamma)$ if the gaugino masses are universal. As explained in detail in Ref.\,\cite{bop}, phenomenologically acceptable models require $\mu M_2>0$.   
In order to satisfy this condition, the gaugino masses are assumed to be generated by a non-zero $F$-term in a 24-dimensional representation of ${\rm SU(5)}\subset {\rm SO(10)}$. This leads to the following pattern of the gaugino masses at the GUT scale \cite{Martin}:
\begin{equation}
\label{gauginoratio}
 (M_1,M_2,M_3) = \left(-\frac{1}{2},-\frac{3}{2},1\right)M_{1/2} \,.
\end{equation}
It should be stressed that even though non-universalities in the gaugino masses are introduced, 
only one free parameter, $M_{1/2}$, governs the gaugino sector.

The soft scalar masses also need to be non-universal to achieve the Yukawa unification 
and light sparticle spectra.\footnote{
In Ref.\,\cite{GogoladzeShafiUn} Yukawa-unified solutions with gaugino masses given in Eq.\,(\ref{gauginoratio}) and universal scalar masses were found. However, the price for the universality in the scalar sector is very heavy spectrum of squarks and gluinos inaccessible at the LHC. Moreover, such solutions predict too small SUSY contribution to the $(g-2)_{\mu}$ to be consistent with the experimental data. In the present paper we demand that the constraint from the $(g-2)_{\mu}$ is satisfied at least at $2\sigma$ level.
}
It is assumed that the $D$-term of the ${\rm U(1)_V}$ (${\rm SO(10)} \supset {\rm SU(5)} \times {\rm U(1)_V}$)
 vector superfield
acquires a vacuum expectation value at the GUT scale.\footnote{
In Ref.\,\cite{Gogoladze_HS_negmu} a model with $\mu<0$ and non-universal gaugino masses
given in Eq.\,(\ref{gauginoratio}) was considered in which sfermion masses are assumed to be universal and ``ad hoc'' Higgs mass splitting is introduced which explicitly breaks SO(10) gauge symmetry.}
The scalar masses, then, can be written as
\begin{align}
\label{scalarDterm}
&m_{H_d}^2=m_{10}^2+2D\,, \nn[4pt]
&m_{H_u}^2=m_{10}^2-2D\,, \nn[4pt]
&m_{Q,U,E}^2=m_{16}^2+D\,, \nn[4pt]
&m_{D,L}^2=m_{16}^2-3D\,. 
\end{align} 
Such a contribution generically arises in the effective theory below the GUT scale as a consequence of spontaneous breakdown of the ${\rm U(1)_V}$ at the GUT scale \cite{Dterm}. Finally, it is assumed that the soft trilinear couplings have a universal value $A_0$ at the GUT scale\footnote{
This assumption requires the existence of a singlet F-term which dominates the soft trilinear couplings but gives a subdominant contribution to the gaugino masses.  The phenomenology of models without the singlet F-term (and with the non-universal A-terms) will be investigated 
in a separate article.
}.

It was shown in Ref.\,\cite{Murayama} that for the universal gaugino masses  the top-bottom-tau Yukawa unification can be realized only if $D>0$ and $m_{10}>m_{16}$. The same conditions 
are preferred 
in the SO(10) model studied in Ref.\,\cite{bop} where the gaugino masses 
are given in Eq.\,(\ref{gauginoratio}). 
Moreover, it was noted in Ref.\,\cite{bop} that the Yukawa unification consistent with the REWSB requires specific correlations between parameters $D$, $m_{10}$ and $A_0$. Provided that these correlations hold, the top-bottom-tau Yukawa unification can be realized for a very wide range of $M_{1/2}$ and $m_{16}$. 
Non-trivial constraints on the model 
are obtained from
the BR$(b\to s \gamma)$, the $(g-2)_{\mu}$ 
and the upper bound on the relic abundance of the lightest neutralinos, $\Omega_{\rm DM}h^2$. Demanding that the predictions of the model for these observables are consistent with the experimental data, the assumption of Yukawa unification leads to rather definite prediction for the MSSM spectrum. This is mainly due to the tension between the BR$(b\to s \gamma)$ and the $(g-2)_{\mu}$: the former prefers heavy SUSY spectra while the latter the light ones. 
On top of that, the WMAP bound \cite{WMAP7} on the $\Omega_{\rm DM}h^2$ can be satisfied only in some corners of the parameter space where various mechanisms which enhance the annihilation cross-section of the LSP are at work. It was found in Ref.\,\cite{bop} that in the part of the parameter space where the Yukawa unification is consistent simultaneously with the BR$(b\to s \gamma)$ and the $(g-2)_{\mu}$, the gluino mass is predicted to have any value below about 1550 GeV. 
However, in the region of the parameter space where the gluino mass is below about 500 GeV and between about 700 and 900 GeV, there is no efficient LSP annihilation mechanism that could reduce the thermal relic abundance of the neutralinos to acceptable values. 

For the gluino mass between about 500 and 700 GeV WMAP bound on the $\Omega_{\rm DM}h^2$ can be satisfied due to resonant annihilations of the LSP through the $Z$ boson or the lightest CP-even Higgs exchange. Throughout this paper we will refer to the region of the parameter space with the efficient LSP annihilation via $Z$ or $h^0$ as the light gluino region. In light of the recent LHC data it is not clear whether the model points with such a light gluino are still allowed. In Ref.\,\cite{ATLAS0lepton} ATLAS collaboration presented interpretation of the 0-lepton search in the simplified model in which all the sparticles except for the gluino, degenerate squarks of the first and second generations and the massless neutralino are decoupled. 
In such a model gluino with mass below 700 GeV has been excluded at 95\% C.L. for the squark masses up to 2 TeV, with the limit increasing to 1075 GeV for equal mass squarks and gluinos. In the light gluino region of the SO(10) model of Ref.\,\cite{bop}, the masses of $\tilde{u}_R$, $\tilde{u}_L$ and $\tilde{d}_L$ (which are degenerate to large extent) are in the range 1.1-1.6 TeV, while the mass of $\tilde{d}_R$ (which is always lighter than the other first generation squarks due to the $D$-term splitting of soft scalar masses) is between about 900 and 1300 GeV. Applying limits on the gluino and squark masses in the simplified model to that SO(10) model one would exclude the light gluino region.
However, the limits on the gluino and squarks masses in the simplified model described above should be treated as an useful indication to what extent the gluino and squark masses can be constrained with the available LHC data. 
For realistic models, these limits will be typically weaker. One of the main purposes of the present paper is to check whether the light gluino region of the SO(10) model of Ref.\,\cite{bop} is still allowed by the recent LHC data.

There are at least two reasons why the limits from the 0-lepton search on the squark and gluino masses in the SO(10) model of \cite{bop} may be substantially weaker than in the simplified model. 
First, the winos and Higgsinos are always lighter than the gluino at the EW scale.\footnote{
$M_3 \sim -2 M_2 \sim -12 M_1$ at the EW scale.  
}
Therefore, the gluino usually decays to the chargino or neutralino other than the bino-like LSP, 
unlike in the simplified model where the gluino decays to $q\ov{q}\tilde{\chi}^0_1$ through an off-shell squark. Second, in some part of the parameter space the right-handed sbottom is lighter than the gluino. For such model points gluino decays exclusively to $\tilde{b}_1$ (associated  with a b-jet) which subsequently decays most of the times to the Higgsino-like neutralino or chargino unless these decays are kinematically inaccessible. 
Longer gluino decay chains result on average in smaller $\pt$ of individual jets and such events are less likely to pass the selection cuts.

In the second region with the gluino mass between about 900 and 1550 GeV, three mechanisms enhancing the LSP annihilations may be at work: t-channel exchange of the right-handed sbottom, resonant annihilation through the CP-odd Higgs, $A^0$, and coannihilations with the stau. The first aforementioned mechanism requires the sbottom to be lighter than about 200 GeV. 
The Yukawa-unified solutions with such a light sbottom have been found in the numerical scan performed in Ref.\,\cite{bop} but only in the part of the parameter space where the LSP mass is below 100 GeV. Such solutions are excluded at 95\% C.L. by the D0 search at the Tevatron \cite{D0sbottom}. 
The Yukawa-unified solutions, in which 
the LSP annihilation via the $A^0$ exchange dominates the total annihilation cross-section,
were ruled out very recently 
by the LHC direct $A^0\to \tau\tau$ searches.
This follows from the fact that 
the LSP annihilation via the $A^0$ exchange in the SO(10) model of Ref.\,\cite{bop} can be large enough
only when $m_{A^0}\lesssim 400$ GeV and $\tan\beta$ between 45 and 50,
while the ATLAS \cite{ATLASmA} and CMS \cite{CMSmA} analyses excluded $m_{A^0}$ below 400 (450) GeV for $\tan\beta\approx$ 45 (50) at 95\% C.L.. 
The limits on the $(m_{A^0},\tan\beta)$ plane do not depend significantly on the other MSSM parameters \cite{mAlimitdependence}, and these limits are also applicable to that SO(10) model in a rather good approximation. 

The regions of the parameter space where the WMAP bound is satisfied due to 
the stau coannihilation are not constrained by any experiment so far. 
The values of the gluino mass found in  the stau-coannihilation region are between about 1000 and 1550 GeV, the masses of the $\tilde{u}_L$ and the $\tilde{d}_L$ are in the range 1.3-1.8 TeV, while the $\tilde{u}_R$ ($\tilde{d}_R$) are lighter by about 100 (300) GeV than  the left-handed squarks.
In contrast to the light gluino mass region, in the stau-coannihilation region, the gluino is always heavier than all the third generation squarks and often heavier than the $\tilde{d}_R$. 
Most importantly, the $\tilde{b}_1$ is always lighter than the gluino by at least 500 GeV which implies that the sbottom production is comparable or even more significant than the gluino and first generation squarks production.   
Notice that the spectra corresponding to the stau-coannihilation region are too heavy to be excluded even if one applies the limits found in the simplified model with the gluino, the LSP and only the first generation squarks (based on the 0-lepton search) and the simplified model with the gluino, the LSP and the lighter sbottom (based on the b-jet search) which typically overestimate the true exclusion limits. 
Nevertheless, we include this region in our analysis to confirm this expectation.

There is also a part of the parameter space where both the $A^0$ exchange and the stau-coannihilation contribute to the LSP annihilation cross-section at similar rate and the WMAP bound is satisfied.
In this region, the required $A^0$ mass has not been excluded by the current LHC searches.
However, we should comment that such solutions require $m_{A^0}$ to be smaller than about 480 GeV, 
and those solutions may be excluded when the whole LHC data accumulated in 2011 
are analysed. 
The gluino and squark masses
corresponding to such solutions are similar to those 
for the stau-coannihilation region.

We should also comment on the fact that the only Yukawa-unified solutions corresponding to the gluino mass in the range $900-1000$ GeV are those with a very light sbottom or $m_{A^0}\lesssim400$. As we pointed out above, the former are excluded by the Tevatron while the latter are excluded by the LHC. This is an interesting example of a case in which parameter $M_{1/2}$ and thus the gluino mass can be constrained not because of excessive production of the gluinos in the $pp$ collisions but as a consequence of some correlations between the MSSM parameters in a specific subclass of MSSM models.

\begin{table}[tbp]
\begin{center}
\begin{tabular}{c|c|c|c|c|c}
\hline
Signal Region & $\geqslant$ 2 jets & $\geqslant$ 3 jets & $\geqslant$ 4 jets low & $\geqslant$ 4 jets high & High mass\\
\hline\hline
$\Etmiss$  &  $>130$ & $>130$ & $>130$ & $>130$ & $>130$\\
Leading jet $\pt$  & $>130$ & $>130$ & $>130$ & $>130$ & $>130$\\
Second jet $\pt$ & $>40$ & $>40$ & $>40$ & $>40$ & $>80$\\
Third jet $\pt$  & - & $>40$ & $>40$ & $>40$ & $>80$\\
Fourth jet $\pt$  & - & - & $>40$ & $>40$ & $>80$\\
$\Delta\phi_{\rm min}$ & $>0.4$ & $>0.4$ & $>0.4$ & $>0.4$ & $>0.4$\\
$\Etmiss/\meff$ & $>0.3$ & $>0.25$ & $>0.25$ & $>0.25$ & $>0.2$\\
$\meff$  & $>1000$ & $>1000$ & $>500$ & $>1000$ & $>1100$\\
\hline\hline
SM Background & 62.4$\pm$4.4$\pm$9.3 & 54.9$\pm$3.9$\pm$7.1 & 1015$\pm$41$\pm$144 & 33.9$\pm$2.9$\pm$6.2 & 13.1$\pm$1.9$\pm2.5$\\
Data & 58 & 59 & 1118 & 40 & 18 \\
\hline 
\end{tabular}
\caption{\small{
The cuts used to define the signal regions of the ATLAS 0-lepton analysis \cite{ATLAS0lepton}. All the dimensionful quantities are given in GeV. 
The $\Delta\phi_{\rm min}$ is the minimum azimuthal angle between $\ptmiss$ and $\ptvec$ of the three leading jets. The number of events observed by ATLAS in each signal region, as well as the expected Standard Model background are also displayed. The first uncertainty represents the statistical uncertainty, while the second is the systematic one.
}}
\label{tab_0lepcuts}
\end{center}
\end{table} 

Another interesting feature of the SO(10) model is the fact that the lightest CP-even Higgs is predicted to be below about 114.5 GeV in the region of the parameter space where the $(g-2)_{\mu}$ and the $b\to s \gamma$ are simultaneously satisfied. This is dangerously close to the LEP2 bound \cite{LEPbound} but taking into account the fact that the uncertainty of this prediction is about 3 GeV \cite{Allanach_higgs}
this region of the parameter space is still allowed by the experimental data.  We should also emphasise that the lightest Higgs can be much heavier if the constraint from the $(g-2)_{\mu}$ is neglected. In particular, the mass of the Higgs can be about 125 GeV, as suggested by the excess observed recently by ATLAS \cite{ATLASHiggs5fb} and CMS \cite{CMSHiggs5fb}, but the corresponding masses of the squarks and the gluino are too heavy to be discovered at the LHC with $\sqrt{s}=7$ TeV.

%%%%%%%%%%%%%%%%%%%%%%%%%%%%%%%%%%%%%%%%%%%%%%%%%%%%%%%%%%%%%%%%%%%%%%
%%%%%%%%%%%%%%%%%%%%%%%%%%%%%%%%%%%%%%%%%%%%%%%%%%%%%%%%%%%%%%%%%%%%%%

%%%%%%%%%%%%%%%%%%%%%%%%%%%%%%%%%%%%%%%%%%%%%%%%%%%%%%%%%%%%%%%%%%%%%%
%%%%%%%%%%%%%%%%%%%%%%%%%%%%%%%%%%%%%%%%%%%%%%%%%%%%%%%%%%%%%%%%%%%%%%

\section{ATLAS searches for squarks and gluinos}
\label{sec_ATLAS}

In order to investigate the LHC constraints on the SO(10) model we focus in this paper on 
the SUSY searches performed by ATLAS, 
although one can expect that the limits from the CMS searches are comparable. The 0-lepton search performed by ATLAS \cite{ATLAS0lepton} with 1.04 fb$^{-1}$ of data defines five signal regions with at least two, three or four jets. The main kinematic variables used in that analysis are the effective mass, $\meff$, and the magnitude of the missing transverse momentum $\Etmiss\equiv|\ptmiss|$. 
The $\meff$ is defined as the sum of $\Etmiss$ and the magnitudes of the transverse momentum, $\pt\equiv|\ptvec|$, of two, three or four leading jets for 
the signal region with at least two, three or four jets, respectively.  
In the ``high mass'' signal region, however, the $\meff$ is defined as the sum of $\Etmiss$ and $\pt$ of all jets with $\pt>40$ GeV.  All the signal regions require high $\pt$ jets and large  $\Etmiss$. The cuts applied in each signal region of the 0-lepton search are summarized in Table \ref{tab_0lepcuts}.

In the SO(10) model of Ref.\,\cite{bop}, the right-handed sbottom is the lightest among the squarks. In some part of the parameter space, the sbottom is even lighter than the gluino. Therefore, one may expect that the sbottoms are significantly produced at the LHC, either directly from the proton-proton collisions or indirectly from the gluino decays. In consequence, a lot of b-jets are expected in a typical final state. That is why we use also the ATLAS b-jets 0-lepton analysis \cite{ATLASbjets} with 0.83 fb$^{-1}$ of data to set constraints on that SO(10) model.  In that analysis, four signal regions have been defined with the requirement of three or more jets and at least one or two b-jets, depending on the signal regions. In the ATLAS b-jet analysis the same kinematic variables as in the 0-lepton search are used. Since one of the main targets of the b-jet 0-lepton analysis is constraining models which predict light sbottoms, the cuts on the $\meff$ are weaker than in the 0-lepton search.  The cuts defining the four signal regions of the b-jet 0-lepton search are summarized in Table \ref{tab_bjetcuts}.

\begin{table}[tbp]
\begin{center}
\begin{tabular}{c|c|c|c|c}
\hline
Signal Region & 3JA & 3JB & 3JC & 3JD\\
\hline\hline
$\Etmiss$  &  $>130$ & $>130$ & $>130$ & $>130$ \\
Leading jet $\pt$  & $>130$ & $>130$ & $>130$ & $>130$ \\
Second jet $\pt$ & $>50$ & $>50$ & $>50$ & $>50$\\
Third jet $\pt$  & $>50$ & $>50$ & $>50$ & $>50$\\
Number of b-jets &  $\geqslant 1$ & $\geqslant 1$ & $\geqslant 2$ & $\geqslant 2$ \\
$\Delta\phi_{\rm min}$ & $>0.4$ & $>0.4$ & $>0.4$ & $>0.4$\\
$\Etmiss/\meff$ & $>0.25$ & $>0.25$ & $>0.25$ & $>0.25$\\
$\meff$  & $>500$ & $>700$ & $>500$ & $>700$\\
\hline\hline
SM Background & $356^{+103}_{-92}$ & $70^{+24}_{-22}$ & $79^{+28}_{-25}$ & $13^{+5.6}_{-5.2}$\\
Data & 361 & 63 & 76 & 12 \\
\hline 
\end{tabular}
\caption{\small{
The cuts used to define the signal regions of the ATLAS b-jets 0-lepton analysis \cite{ATLASbjets}. 
}}
\label{tab_bjetcuts}
\end{center}
\end{table}

Another analysis which may be sensitive to the SO(10) model was released most recently and focuses on search for new phenomena with large jet multiplicities and missing transverse momentum \cite{ATLASmultijets} using 1.34 fb$^{-1}$ of data.  Throughout this paper we will refer to that analysis as the multijets analysis. In that analysis events with leptons in the final state are discarded. Four different signal regions have been defined with the minimum number of jets varying between six and eight. For such a large number of jets, large trigger inefficiencies may occur. In order to maintain acceptable trigger efficiency a separation $\Delta R_{jj}>0.6$ between all the jets with $\pt$ above the threshold for a given signal region is required. The cut on $\Etmiss/\sqrt{H_T}$ where $H_T$ is the scalar sum of the transverse momenta of all the jets with $\pt>40$ GeV and $|\eta|<2.8$ implies that also in that search a large missing transverse energy is required. The cuts applied in each signal region of the multijets search are detailed in Table \ref{tab_multijetcuts}.

The primary target of the multijets analysis is to provide increased sensitivity to models with sequential decays to many strongly interacting particles. It therefore fits very well with the purpose of constraining the SO(10) model. The multijets analysis has already been proved to be useful in the context of the CMSSM for which constraints in the region where $m_{1/2}\ll m_0$ from multijets analysis are comparable (or even better if $m_0$ is between about 1 and 1.5 TeV) to the ones from the 0-lepton search \cite{ATLASmultijets}.

\begin{table}[tbp]
\begin{center}
\begin{tabular}{|c|c|c|c|c|}
\hline
Signal Region & 7j55 & 8j55 & 6j80 & 7j80\\
\hline\hline
Jet $\pt$  & \multicolumn{2}{|c|}{$>55$ GeV} & \multicolumn{2}{|c|}{$>80$ GeV} \\
\hline
$\Delta R_{jj}$ & \multicolumn{4}{|c|}{$>0.6$ for any pair of jets}\\
\hline
Number of jets &  $\geqslant 7$ & $\geqslant 8$ & $\geqslant 6$ & $\geqslant 7$ \\
\hline
$\Etmiss/\sqrt{H_T}$ & \multicolumn{4}{|c|}{$>3.5\ {\rm GeV}^{1/2}$}\\
\hline\hline
SM Background & $39\pm9$ & $2.3^{+4.4}_{-0.7}$ & $26\pm6$ & $1.3^{+0.9}_{-0.4}$\\
Data & 45 & 4 & 26 & 3 \\
\hline 
\end{tabular}
\caption{\small{
The cuts used to define the signal regions of the ATLAS multijets analysis \cite{ATLASmultijets}. 
}}
\label{tab_multijetcuts}
\end{center}
\end{table}

For each signal region $i$, the ATLAS papers provide the information on the observed number of events $n_{\rm o}^{(i)}$, the expected number of the Standard Model background events $n_{\rm b}^{(i)}$ together with its error $\sigma_{\rm b}^{(i)}$. These numbers are given in Tables \ref{tab_0lepcuts}, \ref{tab_bjetcuts}, \ref{tab_multijetcuts}.
The agreement between $n_{\rm o}^{(i)}$ and $n_{\rm b}^{(i)}$ can place constraints on 
SUSY models which predict the expected signal events $n_{\rm s}^{(i)}$ with its error $\sigma_{\rm s}^{(i)}$.
In the next section we shall describe our procedure of computing $n_{\rm s}^{(i)}$ and $\sigma_{\rm s}^{(i)}$ and validate it against the ATLAS exclusion plots for the CMSSM (in the case of the 0-lepton \cite{ATLAS0lepton} and the multijets \cite{ATLASmultijets} analyses) and 
the simplified model with the gluino, the sbottom and the lightest neutralino (in the case of the b-jet analysis \cite{ATLASbjets}).

%%%%%%%%%%%%%%%%%%%%%%%%%%%%%%%%%%%%%%%%%%%%%%%%%%%%%%%%%%%%%%%%%%%%%%
%%%%%%%%%%%%%%%%%%%%%%%%%%%%%%%%%%%%%%%%%%%%%%%%%%%%%%%%%%%%%%%%%%%%%%

%%%%%%%%%%%%%%%%%%%%%%%%%%%%%%%%%%%%%%%%%%%%%%%%%%%%%%%%%%%%%%%%%%%%%%
%%%%%%%%%%%%%%%%%%%%%%%%%%%%%%%%%%%%%%%%%%%%%%%%%%%%%%%%%%%%%%%%%%%%%%

\section{Methodology and its validation}
\label{sec_method}

Let us now discuss our methodology of simulating the ATLAS searches. 
We use various publicly available codes.
First, we compute the SUSY spectra with SOFTSUSY \cite{softsusy} whose 
outputs are passed to SUSYHIT \cite{Susyhit} which calculates branching ratios for all SUSY particles. 
We simulate SUSY events using Herwig++ \cite{Herwig} where
the effects of parton shower, hadronization and underling events are taken into account.
10000 events are generated for each model point.
In the simulation we include only production channels with at least one squark or gluino. We do not simulate direct pair productions of the neutralinos, the charginos or the sleptons since they are either negligible or the $\pt$ of jets and the effective mass, $\meff$, corresponding to such events are too small to pass the selection cuts. 
The detector simulation is done using  Delphes \cite{Delphes} with the ATLAS detector card modified according to the ATLAS analyses, where in particular jets are defined using the anti-$k_T$ algorithm with $\Delta R=0.4$ and the b-jet tagging efficiency is set to 50\%. Finally, the total SUSY production cross-section, $\sigma_{\rm NLO}$, is calculated at next-to-leading order with  PROSPINO \cite{Prospino}. 

In order to work out the expected number of SUSY events in the signal region $i$, $n_{\rm s}^{(i)}$, we apply the cuts detailed in Tables \ref{tab_0lepcuts}, \ref{tab_bjetcuts} and \ref{tab_multijetcuts} to find their efficiency $\epsilon$ (i.e. the ratio of the number of events that survived the cuts to the total number of generated events). 
The $n_{\rm s}^{(i)}$ is given by $\sigma_{\rm NLO}\times \epsilon^{(i)} \times {\mathcal L}^{(i)} \times A^{(i)} $, where $A^{(i)}$ is a correction factor which accounts for the impact of additional data cleaning due to electronics failure in one of the calorimeters in the
ATLAS detector during some period of data-taking. We take $A=0.9$ for the b-jet analysis and $A=0.95$ for the 0-lepton and the multijets analyses.

In order to estimate the exclusion limits we mimic the ATLAS analyses to the extent that is allowed by the limited amount of information  provided in the ATLAS publications on those analyses. In particular, the CLs method \cite{CLs} is used which employs a variable $CLs$ defined by:
\begin{equation}
 CLs\equiv\frac{p_{\rm s+b}}{1-p_{\rm b}} \,,
\end{equation}
to set the exclusion limits on a model 
at $(1-CLs)\times 100\%$ confidence level, which means a model is excluded at 95\% C.L. if $CLs<0.05$. In the above definition $p_{\rm s+b}$ is the  p-value of the background plus signal hypothesis while $p_{\rm b}$ is the p-value of the background only hypothesis. Given the information $\vec{\Sigma}^{(i)}=(n_{\rm s}^{(i)},n_{\rm b}^{(i)},\sigma_{\rm s}^{(i)},\sigma_{\rm b}^{(i)})$ for each signal region $i$, the p-values are computed in the following way \cite{AlKhLeWi}: 
\begin{equation}
 p_{\rm s+b}(n_{\rm o}^{(i)})=\sum_{n=0}^{n_{\rm o}^{(i)}} P(n|\vec{\Sigma}^{(i)})\ , \qquad p_{\rm b}(n_{\rm o}^{(i)})=\sum_{n=n_{\rm o}^{(i)}}^{\infty} P(n|\vec{\Sigma}^{(i)}) \,, 
\end{equation}
where $P(n|\vec{\Sigma}^{(i)})$ is the probability of observing $n$ events in  the signal region $i$ which is governed by Poisson distribution with the expectation value for the observed events:
\begin{equation}
 \lambda^{(i)}(\vec{\Sigma}^{(i)},\delta_b,\delta_s)=n_{\rm b}^{(i)} (1+\delta_b \sigma_{\rm b}^{(i)}) + n_{\rm s}^{(i)} (1+\delta_s \sigma_{\rm s}^{(i)}) \,,
\end{equation}
where the impact of the uncertainties on the expected number of events is accounted for by the nuisance parameters $\delta_b$ and $\delta_s$. We have neglected the error on the luminosity since it is negligible as compared to the errors we included. Assuming that the nuisance parameters have Gaussian probability distribution function, the probability of observing $n$ events is given by:
\begin{equation}
P(n|\vec{\Sigma}^{(i)})= \int_{-1/\sigma_s^{(i)}}^{\infty} d\delta_s \int_{-1/\sigma_b^{(i)}}^{\infty}  d\delta_b \frac{e^{-\lambda^{(i)}}(\lambda^{(i)})^n}{n!}e^{-\frac{1}{2}(\delta_s^2+\delta_b^2)} \,.
\end{equation}
The lower limit of the integration is restricted to keep the number of signal and background events independently non-negative.

The ATLAS papers provide the error on the signal cross-section.
It is estimated by varying the renormalization and factorization scales in PROSPINO between half and twice their default value and by considering the PDF uncertainties provided by CTEQ \cite{CTEQ}. The resulting uncertainties on the signal cross-section vary somewhat from point to point in parameter space but it is reasonable to assume that the error on the cross-section for each production channel is constant across the parameter space. In our analysis we assign  uncertainties of 10\%, 20\%, 30\%, 35\% for the $\tilde{q}\tilde{q}^{(*)}$, $\tilde{q}\tilde{g}$, $\tilde{g}\tilde{g}$, $\tilde{t}\tilde{t}^{*}$ and $\tilde{b}\tilde{b}^{*}$  productions, respectively. 

\begin{figure}[t!]
  \begin{center}
    \includegraphics[width=0.5\textwidth]{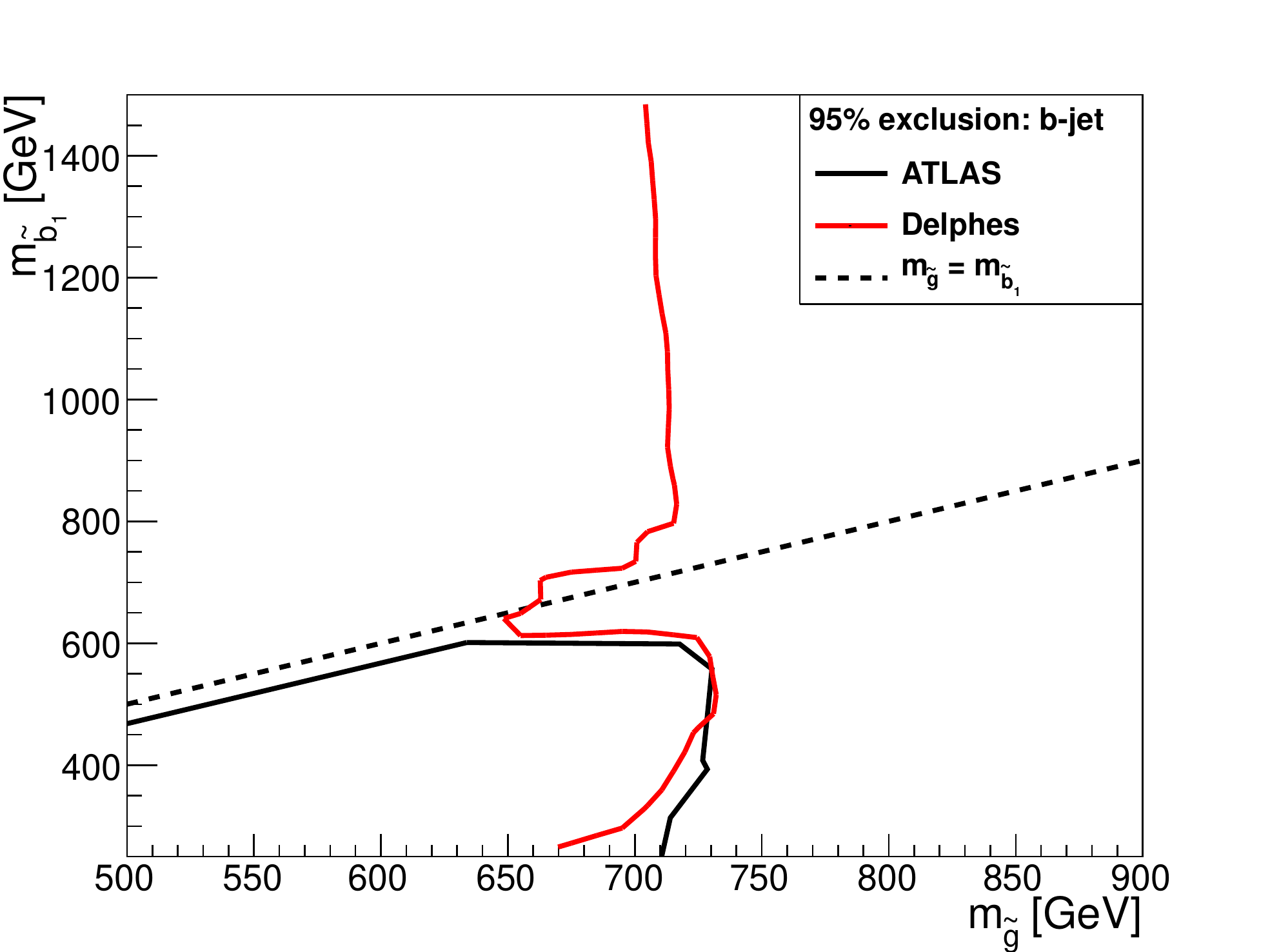}
    \caption{Comparison of our 95\% exclusion contours with those of ATLAS in the simplified model in which only the gluino, the lighter sbottom and the LSP with mass of 60 GeV are in the low energy spectrum, while all the other sparticles are decoupled. In contrast to the ATLAS exclusion contour, our cover also some region of the parameter space where $m_{\tilde{g}}<m_{\tilde{b}_1}+m_b$. }
    \label{bjet_valid}
  \end{center}
\end{figure} 

The ATLAS papers do not provide, however, the details on the other systematic uncertainties on the signal such as the jet energy scale uncertainty or the b-tagging uncertainty. We model these uncertainties by taking a constant value $\sigma_{\rm s'}^{(i)}$ for the signal region $i$. Our estimation of the total systematic uncertainty on the signal is then given by $\sigma_{\rm s}^{(i)}=\sqrt{\sigma_{\rm cross}^2+\sigma_{\rm s'}^{(i)^2}}$. In order to determine $\sigma_{\rm s'}^{(i)}$ we adjust it manually in such a way that our 95\% C.L. exclusion contours match the official ones presented in the ATLAS publications. 

We find that $\sigma_{\rm s'}^{(3JD)}=0.6$ reproduces the official ATLAS exclusion contours from the b-jet analysis \cite{ATLASbjets} in the simplified model containing only the gluino, the sbottom and the lightest neutralino with the mass of 60 GeV reasonably well\footnote{We do note include in our analysis the 3JA, 3JB and 3JC signal regions of the b-jet search because the exclusion plots for the individual signal regions has not been made publicly available so it was not possible to validate our simulation of these signal regions. We do not expect that inclusion of these signal regions would change our conclusions significantly since the 3JD signal region is the most constraining one in the vast majority of the parameter space \cite{ATLASbjets}. }, as seen from Figure \ref{bjet_valid}. In the official ATLAS plot exclusion contours are restricted to the region of the parameter space in which the gluino mass is larger than the sum of the sbottom and bottom masses. In Figure \ref{bjet_valid} we present extended version of that plot to the region of the parameter space where $m_{\tilde{g}}<m_{\tilde{b}_1}+m_b$ and the gluino decays through the off-shell sbottom into $b\ov{b}\tilde{\chi}_1^0$.

We also find that $\sigma_{\rm s'}^{(2j)}=\sigma_{\rm s'}^{(3j)}=\sigma_{\rm s'}^{(4j {\rm high})}=0.3$ and $\sigma_{\rm s'}^{(\rm hm)}=\sigma_{\rm s'}^{(3j)}=\sigma_{\rm s'}^{(4j {\rm low})}=0.1$ give a rather good approximation of the ATLAS exclusion contours for the 0-lepton search, as seen from Figure \ref{0lep_valid} in the Appendix. For the multijets analysis we find the best fit for $\sigma_{\rm s'}^{(7j55)}=\sigma_{\rm s'}^{(8j55)}=\sigma_{\rm s'}^{(6j80)}=\sigma_{\rm s'}^{(7j80)}=0.1$, as seen from Figure \ref{multijets_valid} in the Appendix.

\section{ATLAS constraints on the SO(10) model}
\label{sec_ATLASso10}

We are now ready to present resulting constraints on the SO(10) model from our simulation of the ATLAS analyses. We analysed 977 points (out of which 490 are in the light gluino region) satisfying the top-bottom-tau Yukawa unification at the level of 10\% or better found in a numerical scan performed in Ref.\,\cite{bop}.  All of these points satisfy the following experimental constraints (which were computed with appropriate routines in MicrOmegas \cite{Micromega}):  
\begin{align}
 & 12.7\cdot10^{-10}<\delta a_{\mu}^{SUSY}<44.7\cdot10^{-10}\quad (2\sigma) & \cite{g2} \nn
 & 2.89\cdot10^{-4}<{\rm BR}(b\to s\gamma)<4.21\cdot10^{-4} \quad (2\sigma)  &\cite{bsg} \nn
 & {\rm BR}(B_s\to\mu^+\mu^-)<1.08\cdot10^{-8} &\cite{BsmumuCMSLHCb} \nn
 &  \Omega_{\rm DM}h^2<0.1288 \quad (3\sigma)  &\cite{WMAP7} \nn
 &  m_{h^0}>111.4\,{\rm GeV} \nonumber  & \cite{LEPbound}
\end{align}
Notice that we have included the recent constraint on the BR$(B_s\to\mu^+\mu^-)$ from the combined analysis of CMS and LHCb \cite{BsmumuCMSLHCb}. However, this observable is much less constraining than the lower limit on $m_{A^0}$ which we set in our analysis at 400 GeV independently of $\tan\beta$ (which varies between about 45 and 50 in the region that might be affected by the present constraint on $m_{A^0}$). We use slightly relaxed bound on $m_{A^0}$ due to its (small) model dependence and the theoretical uncertainty in the prediction of $m_{A^0}$. The mass limits on SUSY particles from LEP and the Tevatron including the limit on the sbottom mass discussed in section \ref{sec_YukF24} are also taken into account. Due to the fact that the uncertainty in the prediction of the lightest Higgs boson mass is about $3$ GeV \cite{Allanach_higgs}, slightly relaxed LEP2 bound \cite{LEPbound} is used.\footnote{ Very recently ATLAS \cite{ATLASHiggs5fb} and CMS \cite{CMSHiggs5fb} announced their preliminary results of the SM Higgs searches with 5 fb$^{-1}$ of data. Due to downward fluctuation in the data ATLAS extended the lower limit on the SM Higgs boson mass to 115.5 GeV at 95\% C.L.. We do not take this bound into account because the Higgs boson production cross-section in the MSSM is usually smaller than in the SM due to the negative contribution from a stop loop to the gluon fusion production rate, while ATLAS excluded only $\sigma\gtrsim0.9\sigma_{\rm SM}$ for $m_h\approx115$ GeV.  }    
Even though we require the model points to satisfy only the upper WMAP bound, many of them satisfy also the lower WMAP bound on the dark matter relic density. 

\begin{figure}[t!]
  \begin{center}
     \includegraphics[width=0.6\textwidth]{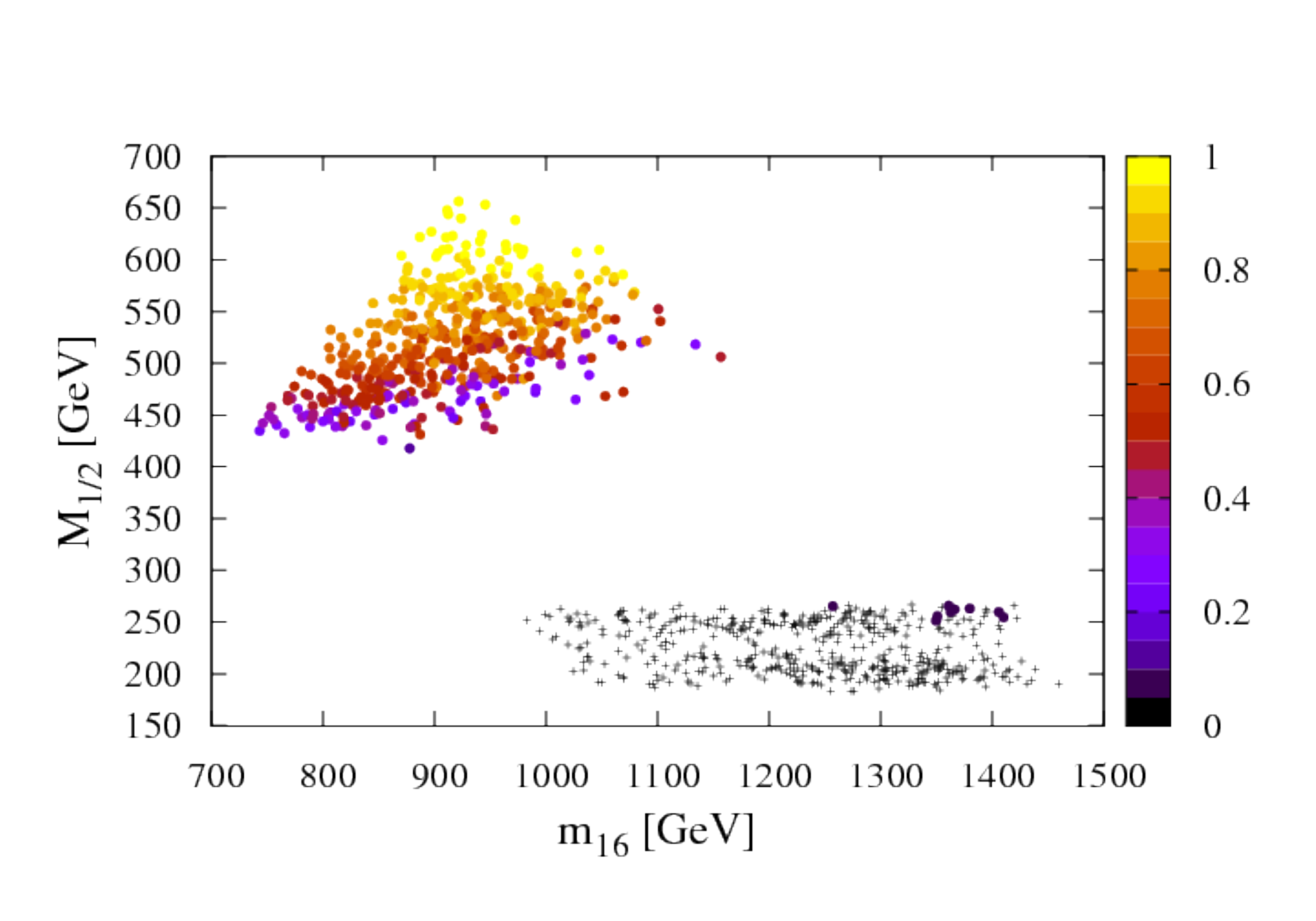}
    \caption{Distribution of the $CLs$ value in the $m_{16}-M_{1/2}$ plane based on the combination of the 0-lepton, the b-jet and the multijets searches. For each point in the plot, the signal region providing the smallest value of the $CLs$ is chosen. The black crosses correspond to the points excluded at 95\% C.L.. }
    \label{fig:m12m16}
  \end{center}
\end{figure}

\begin{figure}[t!]
\begin{center}
\subfigure[b-jet]{\includegraphics[width=0.51\textwidth]{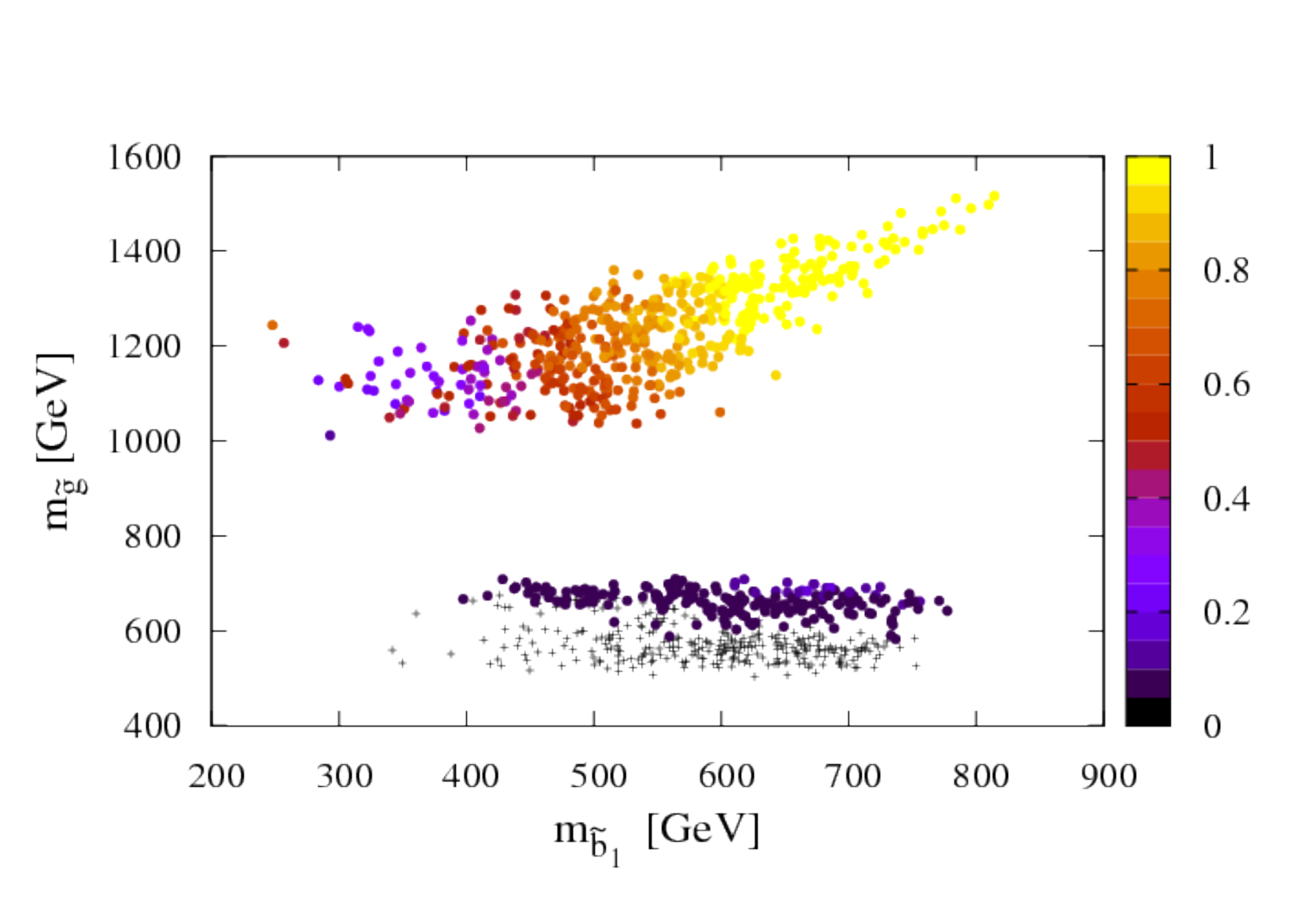}}
\hspace{-17pt}
\subfigure[0-lepton]{\includegraphics[width=0.51\textwidth]{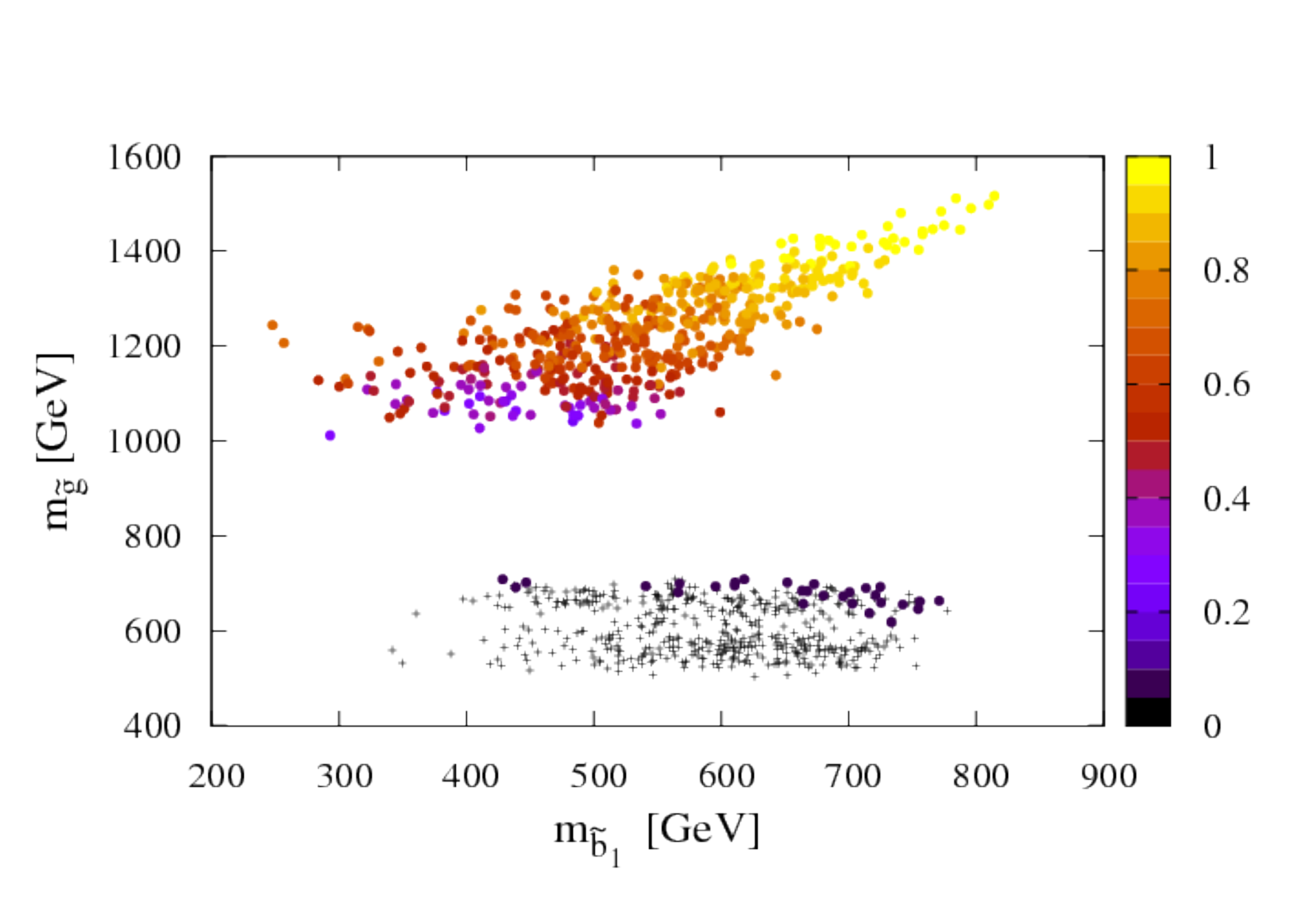}}

\vspace{-25pt}

\subfigure[multijets]{\includegraphics[width=0.51\textwidth]{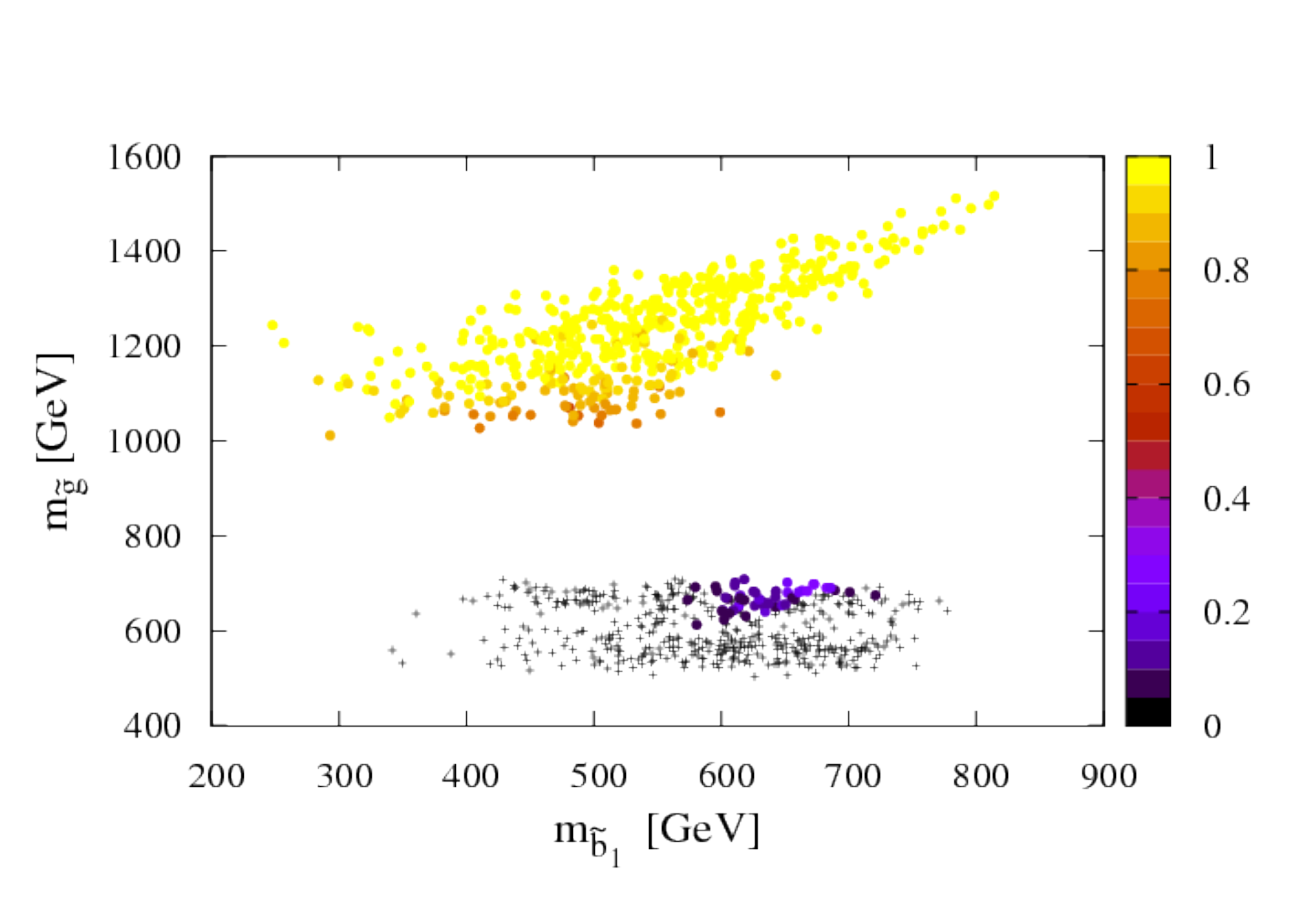}}
\hspace{-17pt}
\subfigure[combined]{\includegraphics[width=0.51\textwidth]{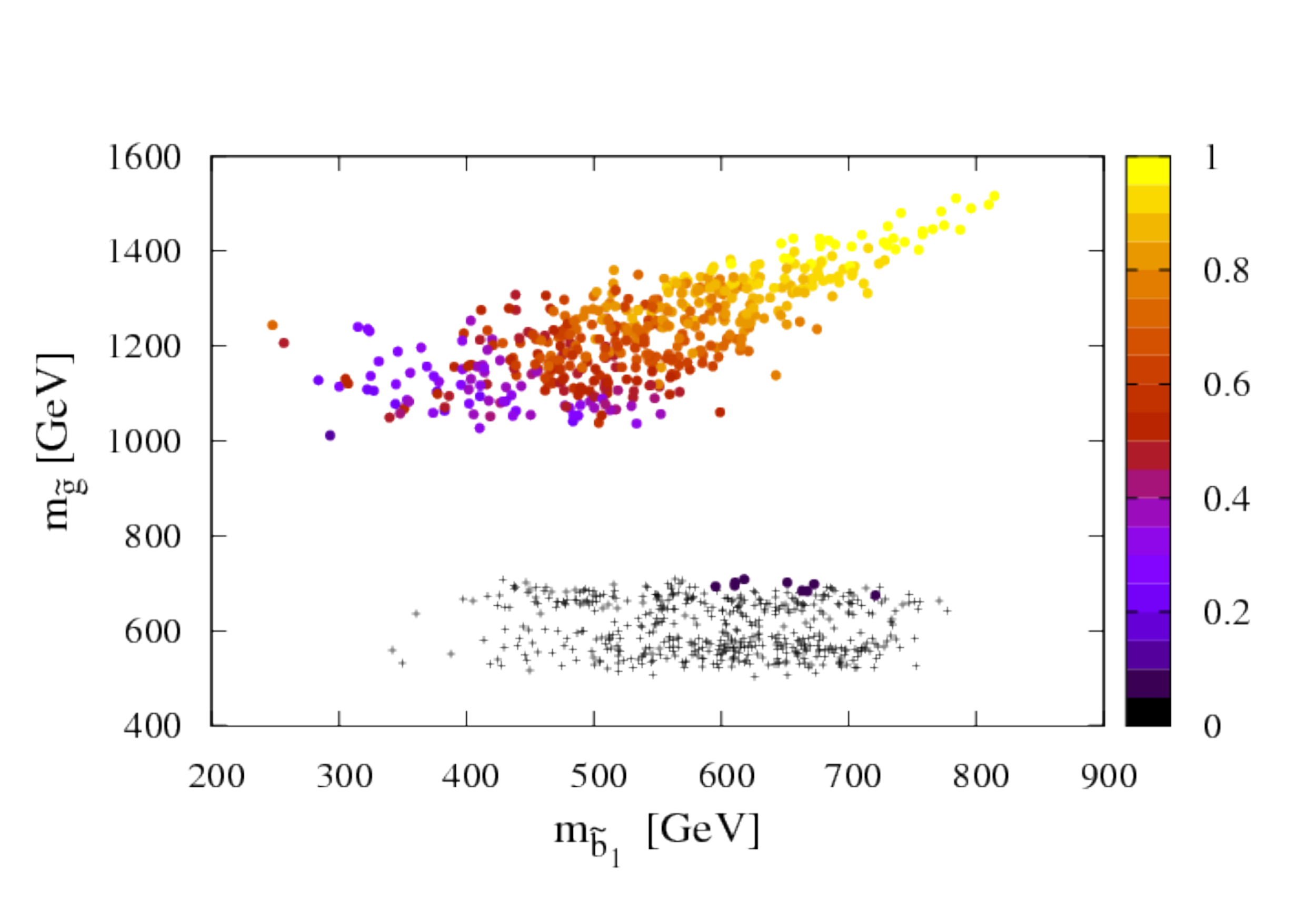}}

\caption{\label{fig:mglmsb} Distribution of the $CLs$ value in the $m_{\tilde{b}_1}-m_{\tilde{g}}$ plane based on: (a) the b-jet search only, (b) the 0-lepton search only, (c) the multijets search only, (d) the combination of all searches. The black crosses correspond to the points excluded at 95\% C.L..
} 
\end{center}
\end{figure}

\begin{figure}[t!]
\begin{center}
\subfigure[b-jet]{\includegraphics[width=0.51\textwidth]{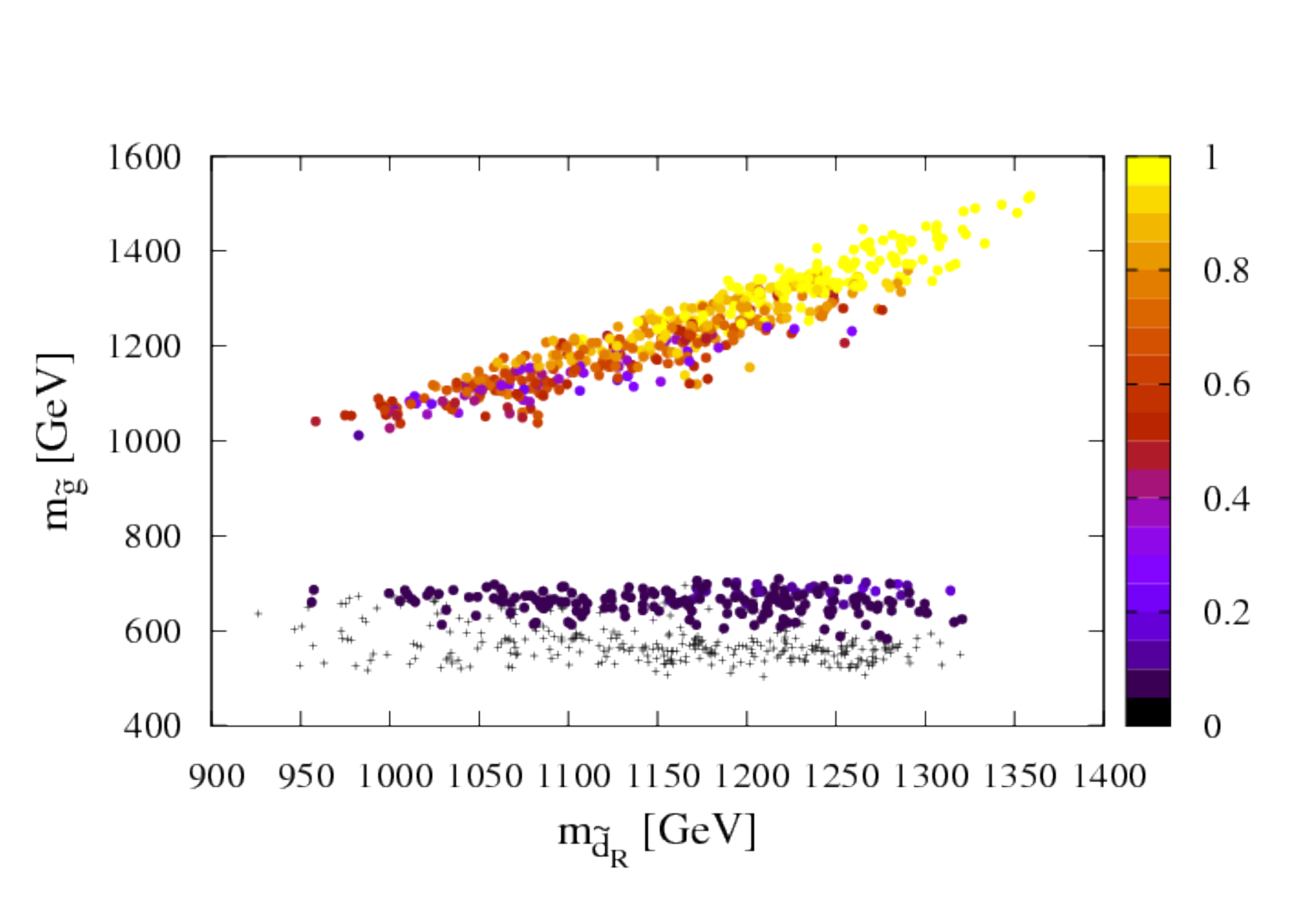}}
\hspace{-17pt}
\subfigure[0-lepton]{\includegraphics[width=0.51\textwidth]{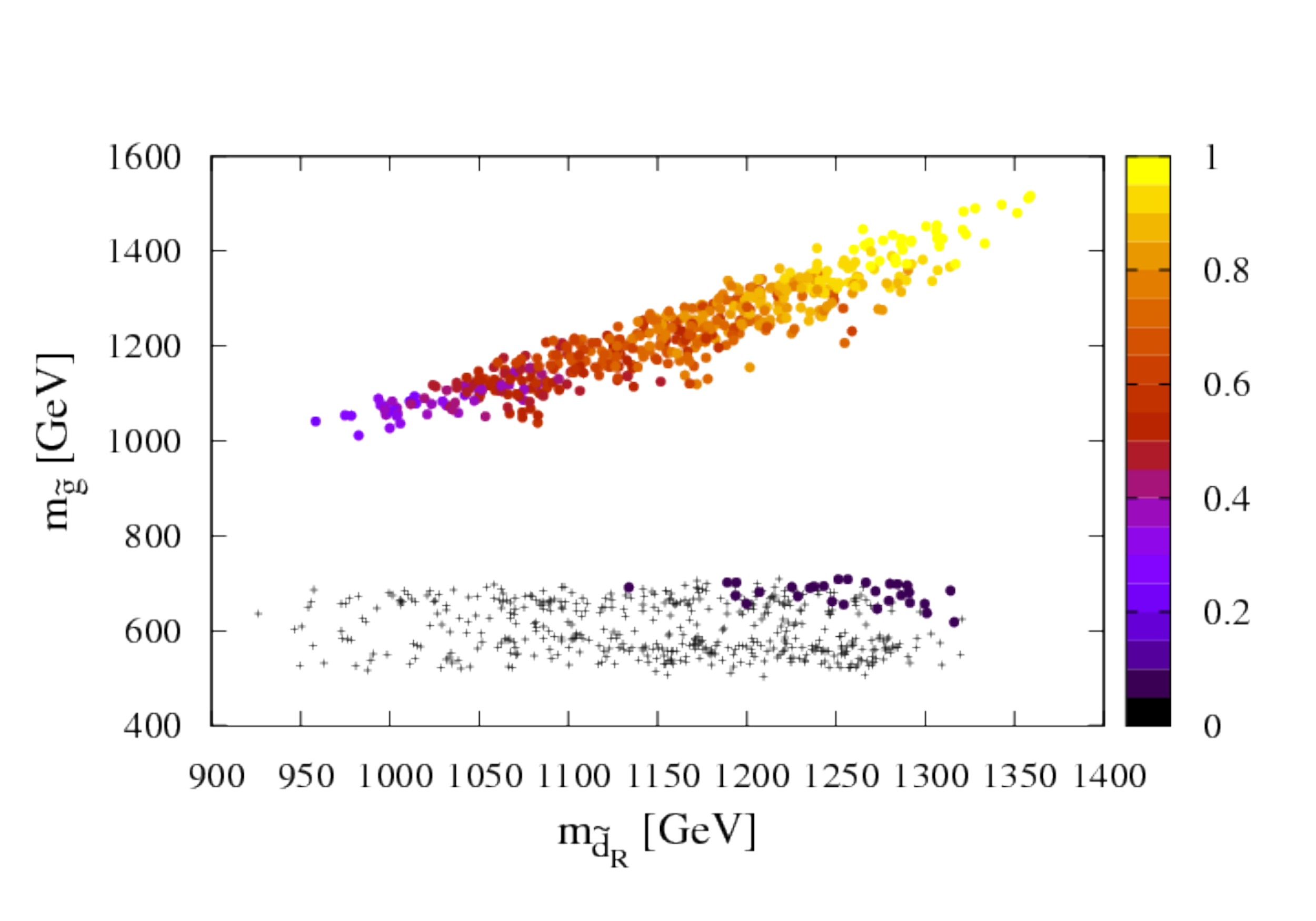}}

\vspace{-25pt}

\subfigure[multijets]{\includegraphics[width=0.51\textwidth]{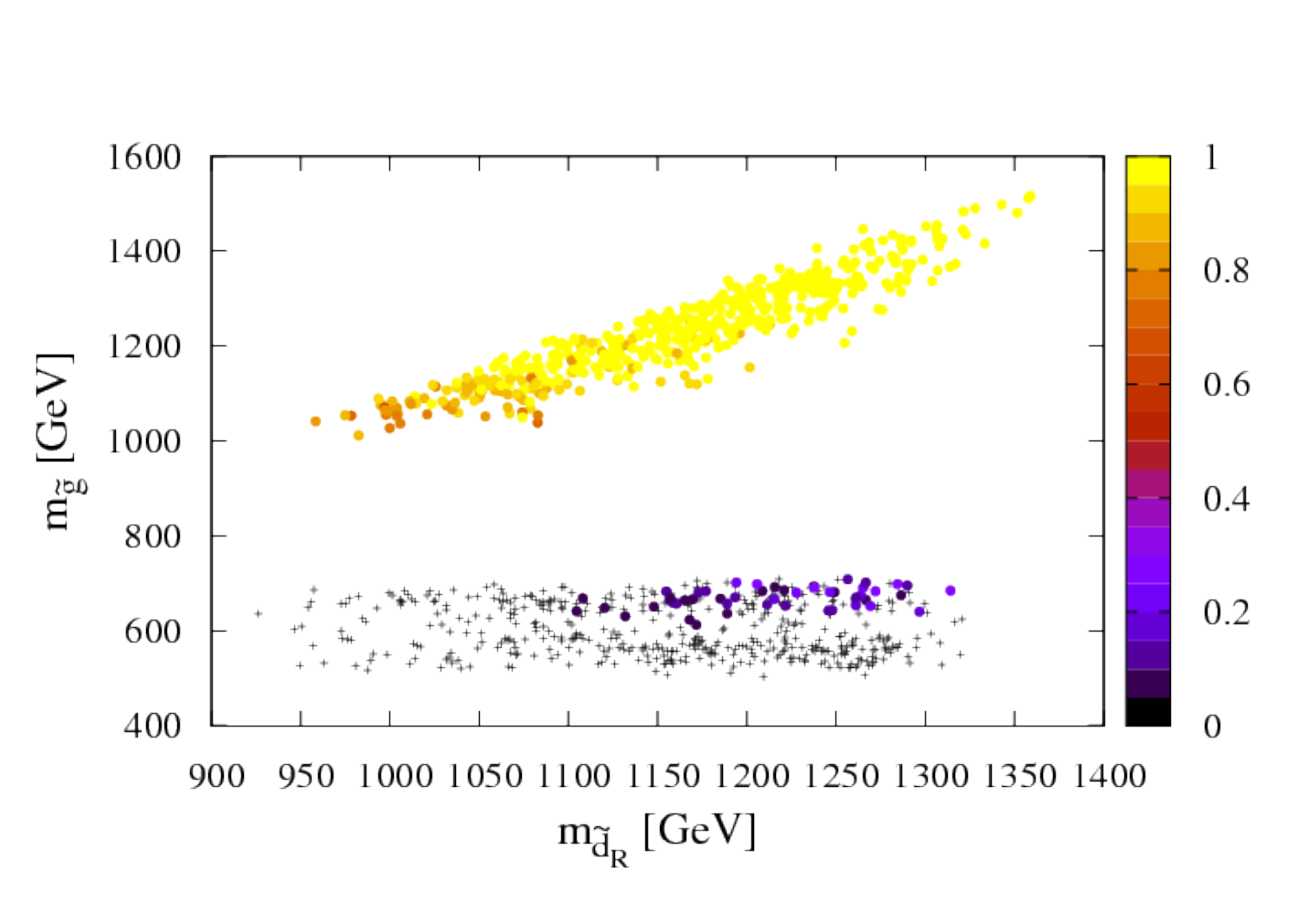}}
\hspace{-17pt}
\subfigure[combined]{\includegraphics[width=0.51\textwidth]{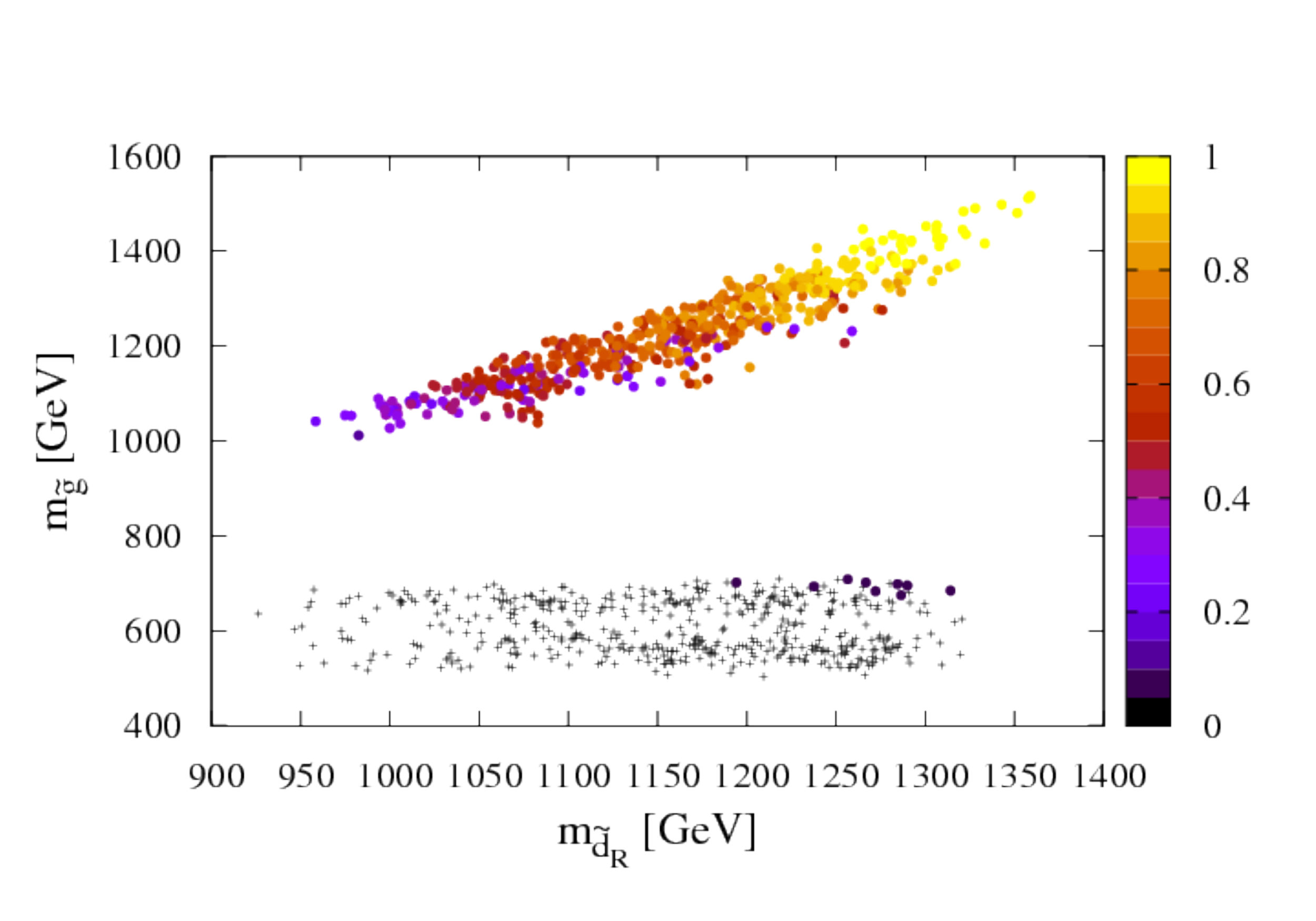}}

\caption{\label{fig:mglmdr} The same as in Figure \ref{fig:mglmsb} but in the $m_{\tilde{d}_R}-m_{\tilde{g}}$ plane. 
} 
\end{center}
\end{figure}
\begin{table}[tbp]
\begin{center}
\begin{tabular}{|c|c|c|c|c|c|c|c|c|c|c|c|c|}
\hline
b-jet & \multicolumn{6}{|c|}{0-lepton} & \multicolumn{5}{|c|}{multijets} & \\
\hline
3JD  & 2j & 3j & 4j low & 4j high & hm & comb & 7j55 & 8j55 & 6j80 & 7j80 & comb & comb  \\
\hline
59.2 & 0 & 8.6 & 0 & 24.5 & 94.1 & 94.1 & 74.3 & 85.9 & 87.3 & 44.1 & 90 & 98.2\\
\hline 
\end{tabular}
\caption{\small{
Percentage of the excluded points at 95\% C.L. in the light gluino region by individual signal regions. ``comb'' in the 0-lepton (multijets) column correspond to a combination of all the signal regions within the 0-lepton (multijets) search, while ``comb'' in the last column corresponds to the combination of all three ATLAS searches.
}}
\label{tab:exclfrac}
\end{center}
\end{table}

The main result of our analysis is that 481 out of 490 points in the light gluino region are excluded at 95\% C.L.. On the other hand, none of the points are excluded in the stau-coannihilation region. This can be seen from Figure \ref{fig:m12m16} where the distribution of the $CLs$ value in the $m_{16}-M_{1/2}$ plane based on the combination of all searches is presented. For each point in this plot, the signal region providing the smallest $CLs$ value is chosen.
The model points depicted by the crosses are excluded at 95\% C.L..

In Figures \ref{fig:mglmsb} and \ref{fig:mglmdr}, the distributions 
of the $CLs$ values  based on the individual searches and their combination 
are shown in the $m_{\tilde{d}_R}-m_{\tilde{g}}$ and $m_{\tilde{b}_1}-m_{\tilde{g}}$ planes, respectively. 
The first thing to note from these Figures is that the 0-lepton and the multijets analyses provide the most stringent constraints on the light gluino region but the b-jet search is only slightly less constraining ruling out by itself about 60\% of points in this region. 
In Table \ref{tab:exclfrac}, fractions of the excluded points in the light gluino region by individual signal regions are shown. In the 0-lepton analysis, essentially only the ``high mass'' region is relevant, while in the multijets analysis all the signal regions are relevant and complement each other. 
We should also emphasise the complementarity between the 0-lepton and the multijets searches. 
While individually these searches constrain the gluino mass to the values above about 620 and 610 GeV, respectively, their combination excludes all the points with the gluino mass below about 675 GeV. 
The multijets analysis is less sensitive to the part of the parameter space where $m_{\tilde g} > m_{\tilde b_1}$ and the mass splitting between the gluino and the sbottom is small. In such a case the gluino to sbottom 
decay produces rather soft jets.    
This implies that passing the cuts with large jet multiplicities in such a case becomes less likely.\footnote{
Note that our setup of the event simulation does not take account of the NLO matrix element correction
with extra hard jets associated with the sparticle production.
Including that contribution may slightly increase the efficiency for the events with soft jets from
SUSY cascade decays.  
In this sense, our analysis provides a conservative limit.   
}
Appearance of the soft jets in the final state affects also the efficiency of the ``high mass'' signal region of the 0-lepton search but in a milder way, so some of the points (but not all) with small mass splitting between the sbottom and the gluino which are not excluded by the multijets analysis are excluded by the 0-lepton search. 
As a result, all the points in the light gluino region which are still allowed are characterized by rather small mass splitting between the gluino and the sbottom - smaller than about 100 GeV, as can be seen from Figure \ref{fig:mglmsb}. Another feature of the points not excluded by any ATLAS search is rather heavy $\tilde{d}_R$. The lower bound on its mass in the light gluino region is found to be about 1.2 TeV, as clearly seen from Figure \ref{fig:mglmdr}.  

It is also interesting to note that all the model points in the light gluino region would be excluded if the ATLAS had not observed some small excess over the background (about $1\sigma$) in the high mass signal region of the 0-lepton analysis (see the last column in Table \ref{tab_0lepcuts}). Therefore, if the excess observed by ATLAS is just an upward fluctuation of the background, and not manifestation of SUSY events, the light gluino region will be excluded very soon.

Even though all of the points in the stau-coannihilation region are consistent with the present LHC data, one can draw some interesting conclusions from our analysis  for this case too. 
First of all, the values of the $CLs$ provided by the b-jet and the 0-lepton search are much smaller than those of the multijets search. This is because in the stau-coannihilation region the mass of the lighter sbottom is typically comparable to the Higgsino mass\footnote{
There are two reasons for this. One is that a requirement of the bottom-tau Yukawa unification positively correlates $|\mu|$ with the gluino mass. The second is that due to large negative contributions to the sbottom mass from $D$-term and the renormalization group effects (coming from a large positive $A_0$ required for consistency with the $b\to s \gamma$ constraint) the sbottom is light (see a detailed discussion in Ref.\,\cite{bop}).}
and the $\tilde{b}_1 \to \tilde{\chi}_1^{-}t$ decay channel is closed. 
The sbottom decaying through this channel may produce as many as five jets if $\tilde{\chi}_1^{-}$ decays subsequently to $\tilde{\chi}_1^{0} W$ so this channel being closed results in a big loss of the cuts efficiency in the multijets analysis. Even if the $\tilde{b}_1 \to \tilde{\chi}_1^{-}t$ decay channel is open, as is the case for a small number of points in the stau-coannihilation region, the $\tilde{\chi}_1^{-}$ usually decays to $\tilde{\nu}_{\tau} \tau$ or $\tilde{\tau}_1\nu_{\tau}$ rather than to $\tilde{\chi}_1^{0} W$ which results in leptons or at least a smaller number of hard jets in the final state making it unlikely that the events would pass the selection cuts used in the multijets analysis.  Moreover, efficiency of the cuts used in the multijets analysis is reduced also by the fact that decays $\tilde{d}_R$ usually decays directly to the LSP producing only single jet. 

The 0-lepton and the b-jet searches set complementary constraints on the stau-coannihilation region. The b-jet search is the most sensitive (i.e. provides the smallest $CLs$ value) to model points with very light sbottom where the total SUSY cross-section is dominated by the sbottom production\footnote{ 
In the final stages of preparing this work new ATLAS analysis was released \cite{ATLASsbottom} in which the limits on the $m_{\tilde{b}_1}-m_{\tilde{\chi}_1^0}$ plane were found that surpass those from D0. In that analysis 2 b-jets are required in the final state and the cuts are optimized in order to maximise the sensitivity for models with the light sbottom. That ATLAS search excludes the model points in the stau-coannihilation region with the sbottom mass below about 350 GeV if BR$(\tilde{b}_1\to \tilde{\chi}_1^0 b)=100\%$ (which is usually the case in this part of the parameter space).
}
while the 0-lepton search provides the best constraints for models with light $\tilde{d}_R$, as seen from Figures \ref{fig:mglmsb} and \ref{fig:mglmdr}.  

\begin{figure}[t!]
\begin{center}
\includegraphics[width=0.51\textwidth]{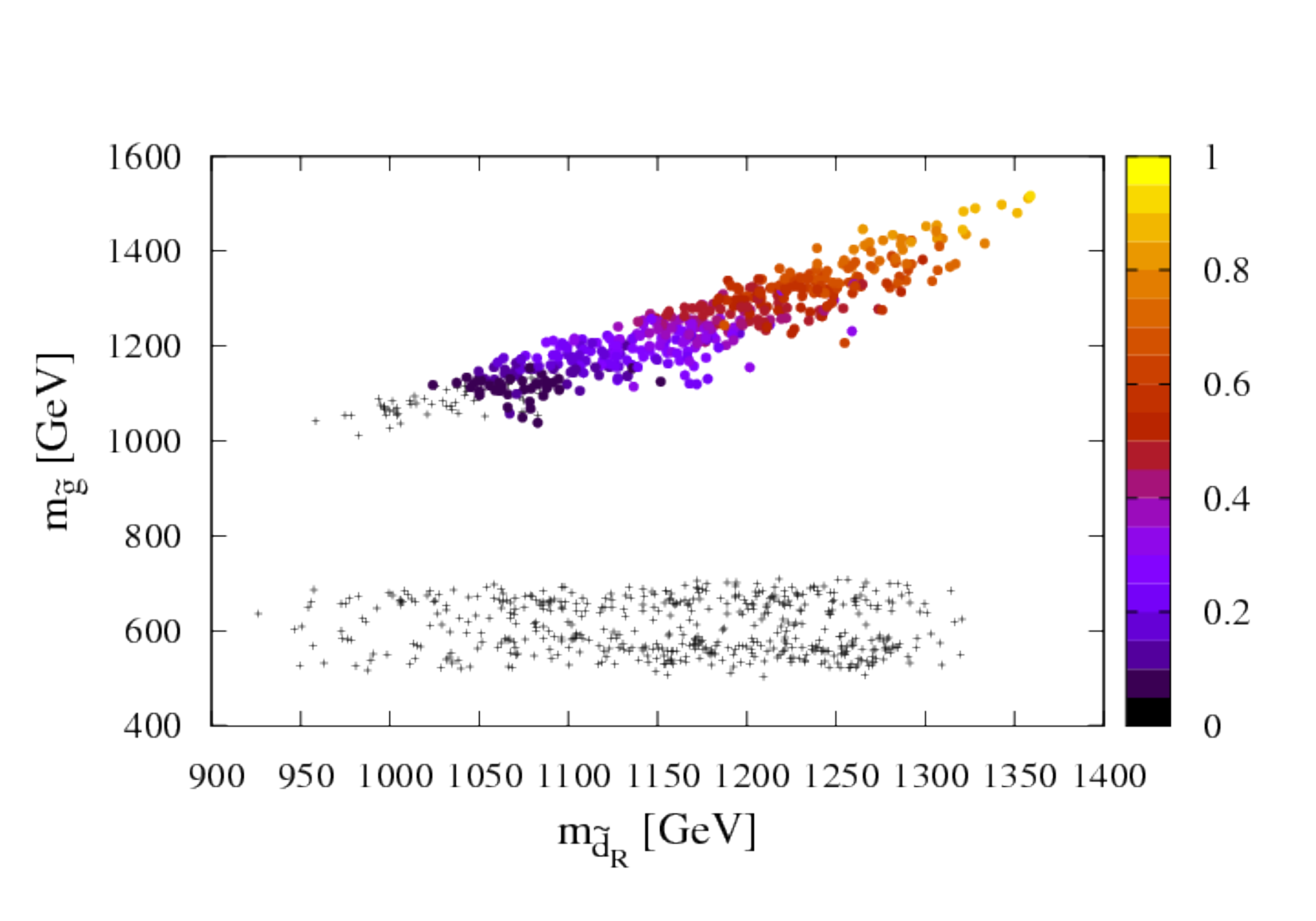}
\hspace{-17pt}
\includegraphics[width=0.51\textwidth]{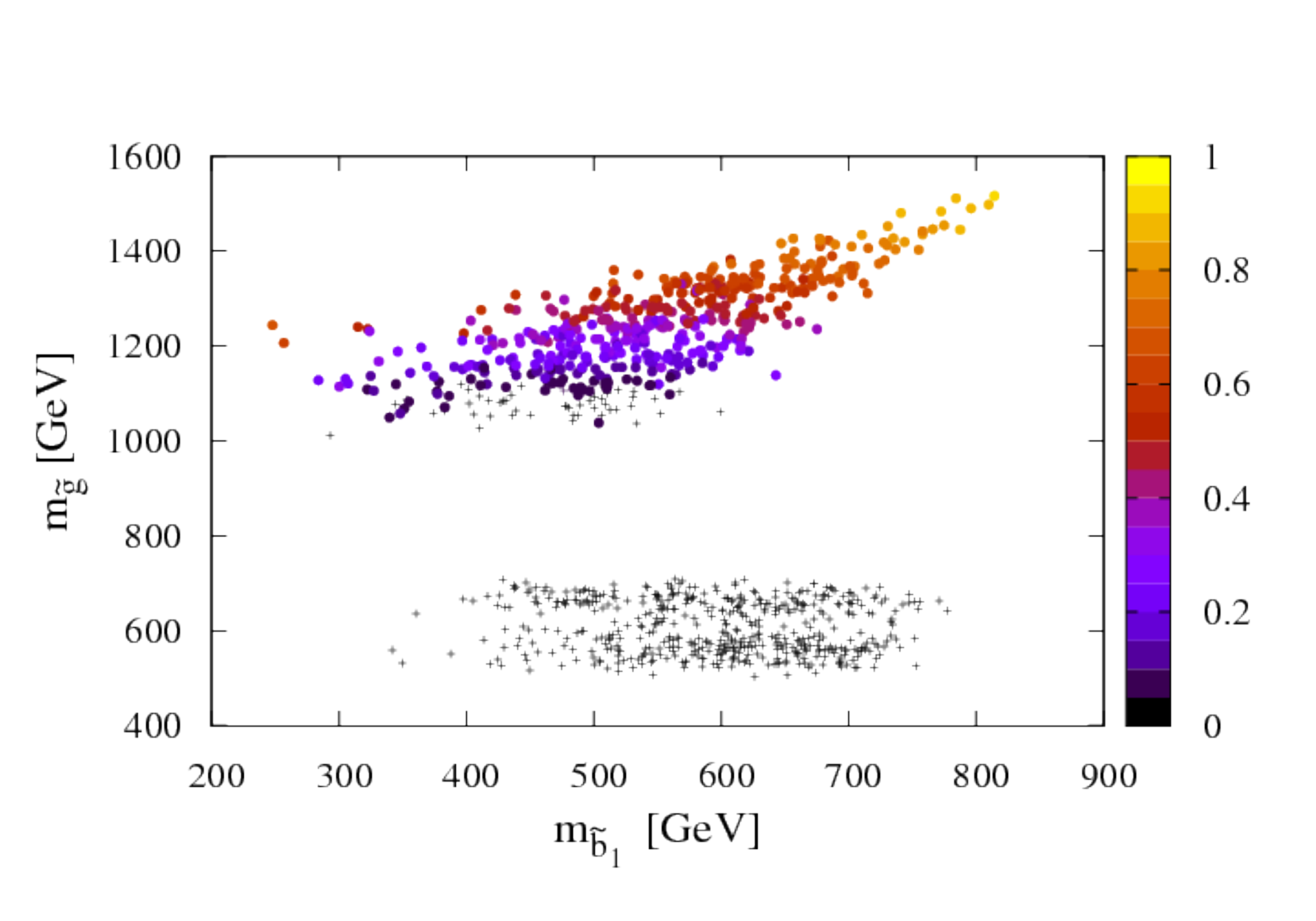}
\caption{\label{fig:0lep_10f} Distribution of the expected $CLs$ value based on the 0-lepton search with 10 fb$^{-1}$ of the integrated luminosity.
} 
\end{center}
\end{figure}

We also estimate the expected constraints on the 
stau-coannihilation region in the SO(10) model with 10 fb$^{-1}$ of the integrated luminosity. The multijets search will not have enough sensitivity to exclude any points from this region, as expected from the above discussion. 
We found that the most promising for constraining the stau-coannihilation region is the 0-lepton search.
It can be seen from Figure \ref{fig:0lep_10f} that the 0-lepton search can probe the gluino masses in the stau-coannihilation region up to 1.1 TeV if the right-handed down squark is lighter than 1 TeV. 
In our estimation of the expected exclusions with 10 fb$^{-1}$, we rescale the statistical errors with the luminosity but keep the relative systematic error constant. This is certainly a conservative assumption.  Thus the reach of the 0-lepton search will be even better.
With this conservative assumption, the b-jet search will not be able to probe the stau-coannihilation region. However, if the systematic uncertainties on the background were to be reduced significantly, for example by data driven-methods of systematic error estimation, and/or the cuts were to be refined 
according to the larger amount of data, then the b-jet search may be complementary to the 0-lepton search, especially in the part of the parameter space with light sbottom.

\section{Summary and conclusions}
\label{sec_concl}

We have investigated the LHC constraints on the recently proposed SUSY SO(10) GUT model \cite{bop} predicting the top-bottom-tau Yukawa unification. In this model the non-universal gaugino masses arising from SO(10) non-singlet $F$-term transforming as ${\bf 24}$ of SU(5) $\subset$ SO(10), $D$-term splitting of the scalar masses and the negative sign of $\mu$ are assumed. The model 
is able to satisfy all experimental constraints including the $(g-2)_{\mu}$, the $b\to s \gamma$ and predict correct thermal relic abundance of the neutralinos. 
In the region of parameter space of the SO(10) model where all phenomenological constraints are fulfilled, relatively light MSSM spectra are predicted and there is a good prospect of discovering SUSY particles at the LHC. We found that this part of the SO(10) model parameter space tightened considerably due to the recent results of the LHC experiments which have not found any sign of SUSY so far, but many Yukawa-unified solutions are still consistent with all experimental data.  

The phenomenologically viable parameter space of this model can be classified 
into three regions in terms of the dominant LSP annihilation process:
the light gluino region (the dominant process is $Z$ boson or light CP-even Higgs exchange), 
the light $A^0$ region (the dominant process is CP-odd Higgs exchange)
and the stau-coannihilation region (the dominant process is stau-coannihilation.).    
We have argued that the light $A^0$ region has already been excluded
by the recent $A^0 \to \tau \tau$ search at the LHC
and we have investigated the impact of the recent direct SUSY searches
on the light gluino and the stau-coannihilation regions in the SO(10) model.

Since the constraints on the gluino and squark masses obtained in the simplified model,
where only the massless neutralino, the gluino and the first/second generation of squarks 
are assumed in a low energy SUSY spectrum,
cannot be directly applied to other models,
we performed a dedicated analysis to assess the impact of the recent direct SUSY searches 
on the SO(10) model.
We focused on three ATLAS analyses: the 0-lepton search with 1.04 fb$^{-1}$ \cite{ATLAS0lepton}, the b-jet search with 0.83 fb$^{-1}$ \cite{ATLASbjets} and the multijets search with 1.34 fb$^{-1}$ \cite{ATLASmultijets} of data. We found the 95\% C.L. lower limit of 675 GeV for the gluino mass, substantially weaker than the one obtained in the simplified model.
This implies that the light gluino region is excluded except for the corner of the parameter 
space where the gluino mass is between 675 GeV and 700 GeV, the right-handed down squark mass is above 1.2 TeV and the mass splitting between the gluino and the lighter sbottom is smaller than about 100 GeV. 
We, however, stress that the light gluino region would be completely excluded in the absence of the slight excess of the events observed in the ``high mass'' signal region of the ATLAS 0-lepton search. 
Therefore, this region will be excluded very soon unless the observed excess is not just an upward fluctuation of the background. 

We also found that the stau-coannihilation region remained unconstrained by the present searches. 
We have estimated the expected exclusion limit on this region assuming the same analyses
but with 10 fb$^{-1}$ of the integrated luminosity at the 7 TeV LHC and found that the 0-lepton search may set the 95\% C.L. lower limit on the gluino mass at 1.1 TeV if the right-handed down squark is not heavier than about 1 TeV.
The b-jet search may also contribute to the constraints on the stau-coannihilation region, especially in the part of parameter space with light sbottom, if the systematic error in that search is reduced and/or the selection cuts are refined.

In conclusion, the SO(10) model of Ref.\,\cite{bop} is still consistent with all available direct and indirect constraints, with good prospects for testing it at the LHC in the near future. This should be contrasted with the Yukawa-unified SO(10) models with the universal gaugino masses where the most favourable part of the parameter space has already been excluded by the LHC experiments. For this reason we believe that the model we have studied in the present paper is the best motivated and the most promising SO(10) GUT model that predicts the top-bottom-tau Yukawa unification. Therefore, it is certainly worth further detailed investigation, both on the theoretical and the experimental side.

%%%%%%%%%%%%%%%%%%%%%%%%%
\section*{Acknowledgments}
%%%%%%%%%%%%%%%%%%%%%%%%%
This work has been partially supported by STFC. We would like to thank A. Barr and the Cambridge SUSY Working group, particularly B. C. Allanach, B. Gripaios and C. Lester, for helpful discussions. MB would also like to thank M. Olechowski for inspiring discussions and S. Pokorski for discussions and comments on the manuscript.
We are indebted to J. Gaunt for reading the first version of this paper and suggesting some refinements to the text.

%%%%%%%%%%%%%%%%%%%%%%%%%%%%%%%%%%%%%%%%%%%%%%%%%%%%%%%%%%
%%%%%%%%%%%%%%%%%%%%%%%%%%%%%%%%%%%%%%%%%%%%%%%%%%%%%%%%%%

\renewcommand{\theequation}{A.\arabic{equation}}
\setcounter{equation}{0}
\renewcommand{\thefigure}{A.\arabic{figure}} 
\setcounter{figure}{0}

\begin{figure}[ht!!]
\begin{center}
\vspace{-10pt}
\subfigure[2 jets]{\includegraphics[width=0.31\textwidth]{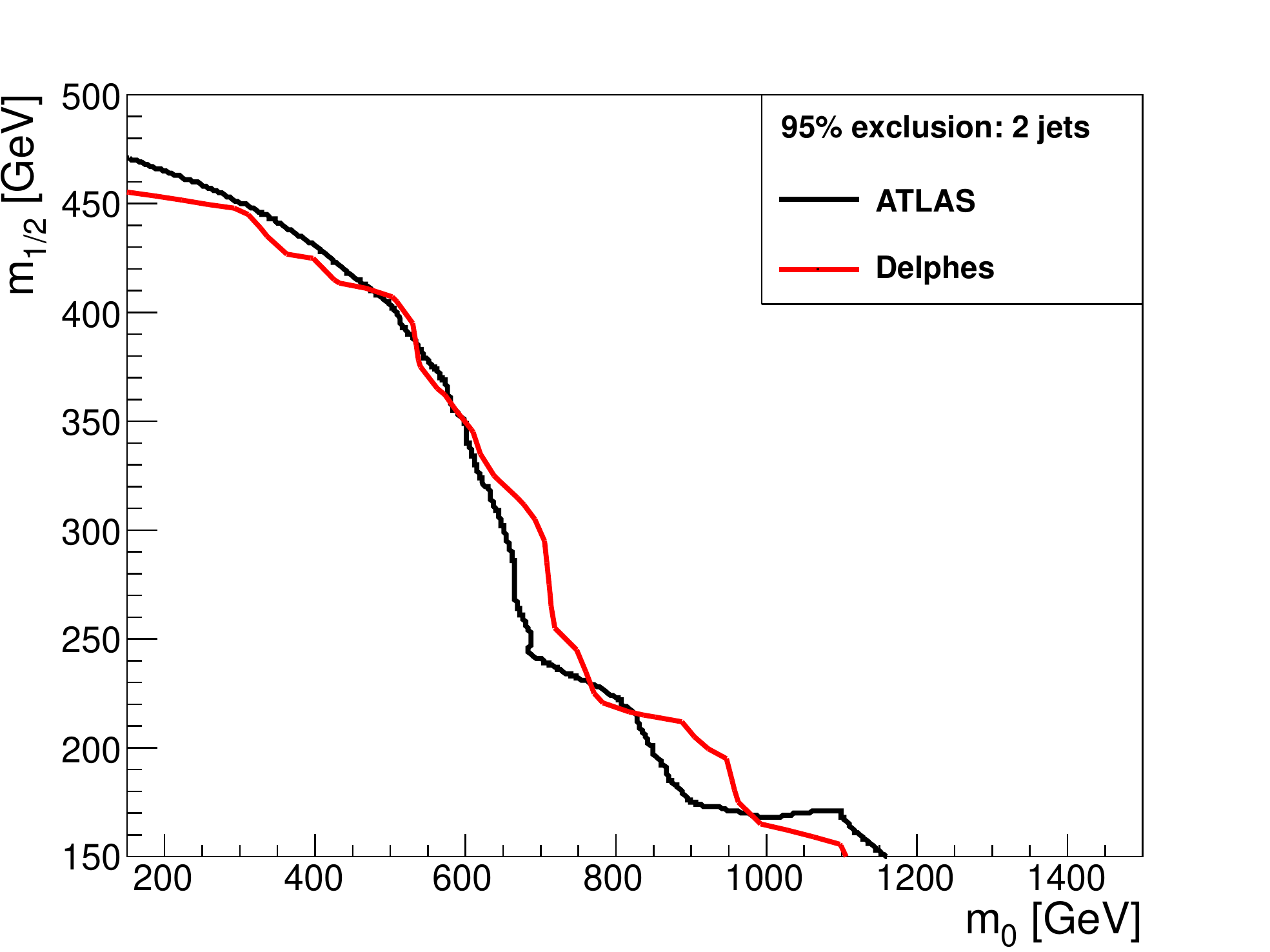}}
\hspace{-15pt}
\subfigure[3 jets]{\includegraphics[width=0.31\textwidth]{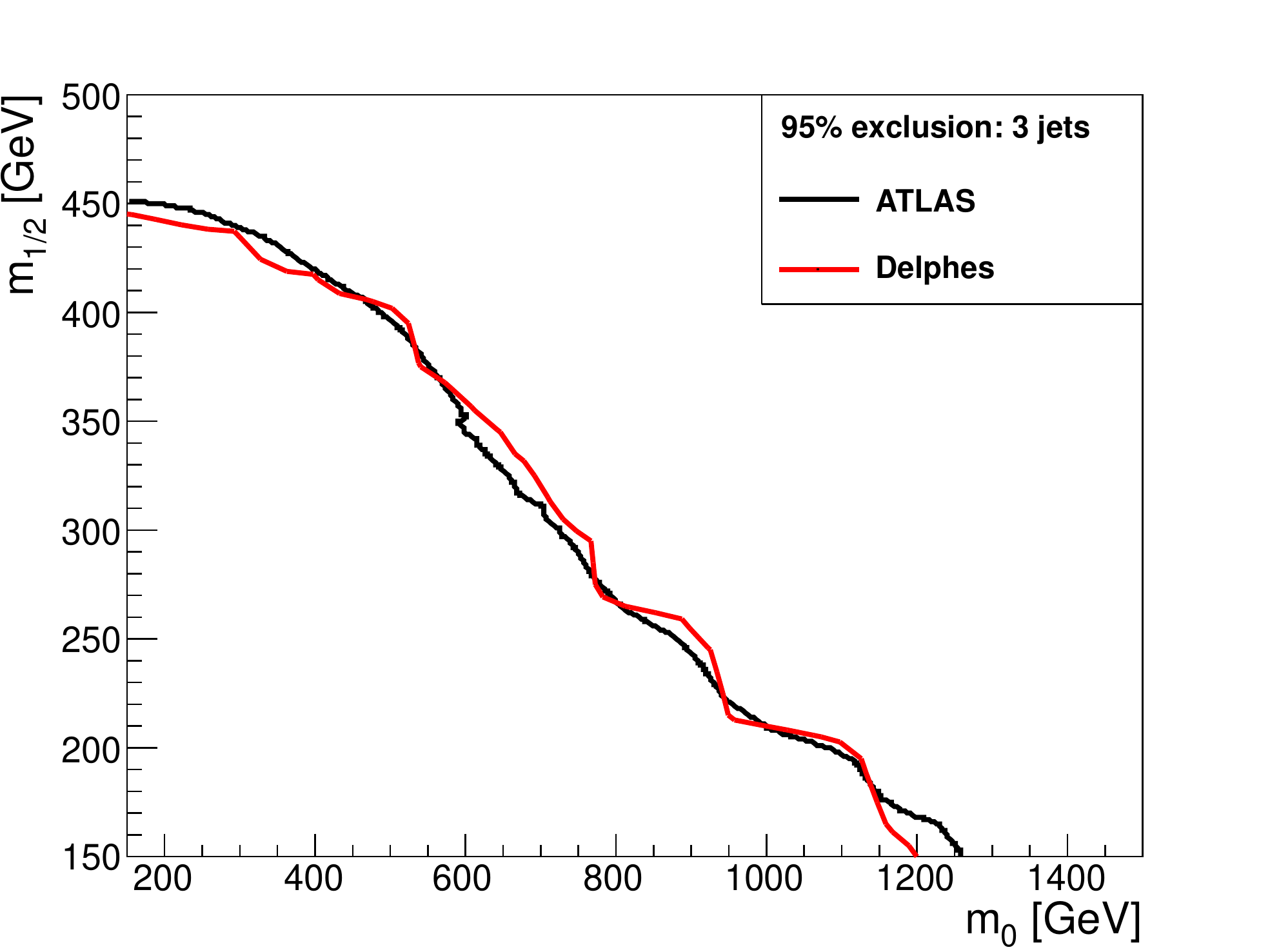}}
\hspace{-15pt}
\subfigure[4 jets low]{\includegraphics[width=0.31\textwidth]{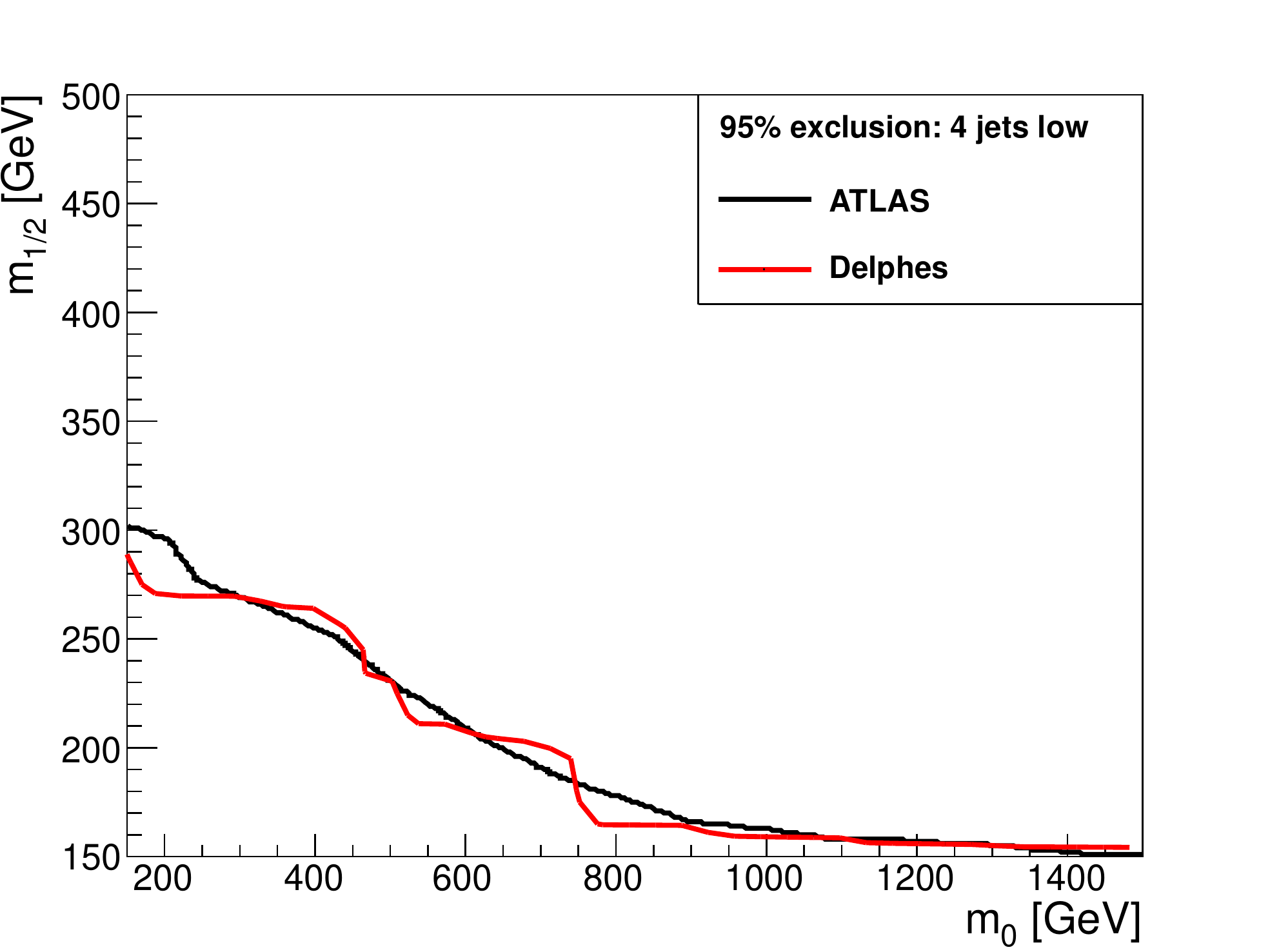}}
\\
\vspace{-10pt}
\subfigure[4 jets high]{\includegraphics[width=0.31\textwidth]{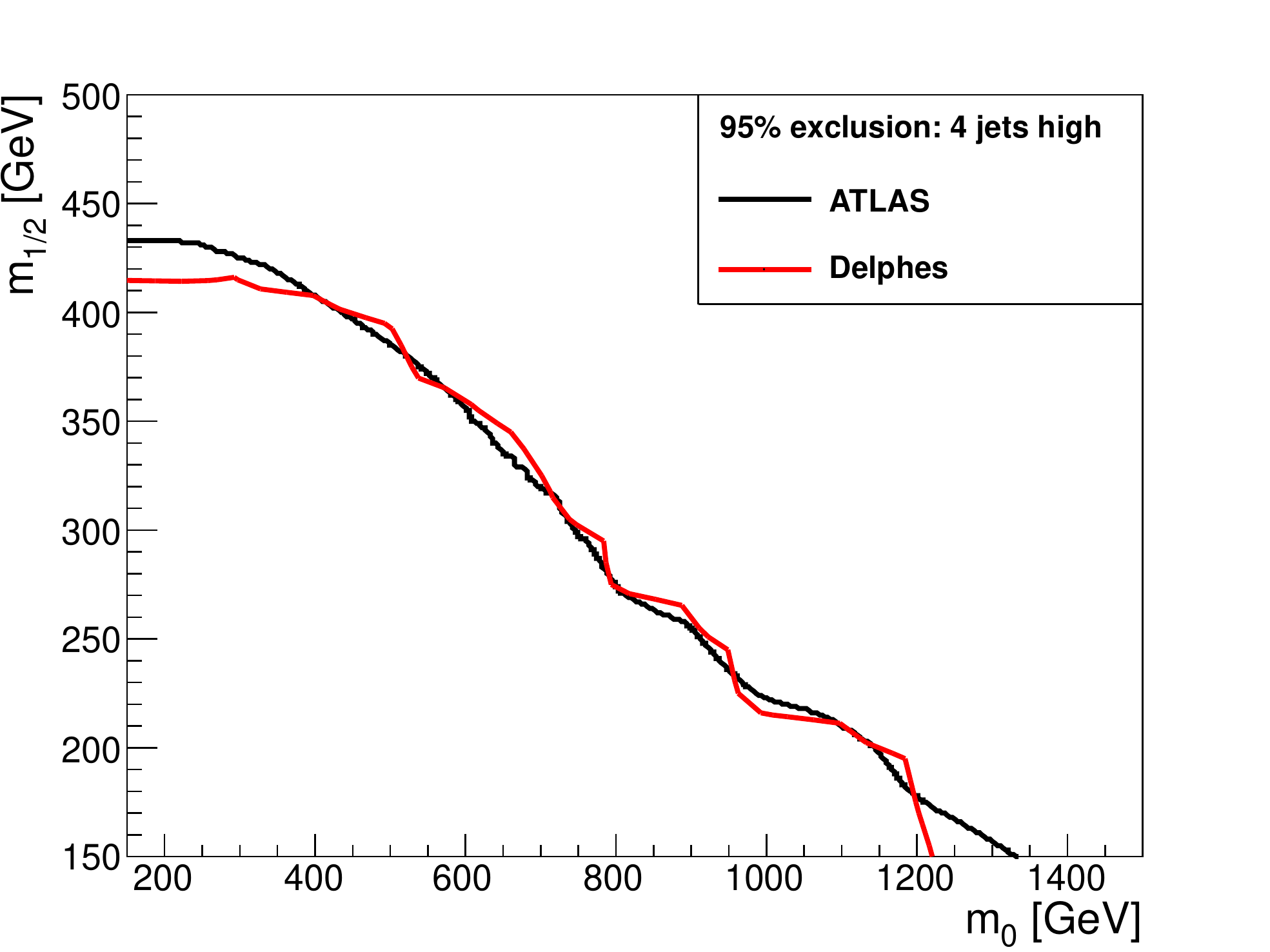}}
\subfigure[high mass]{\includegraphics[width=0.31\textwidth]{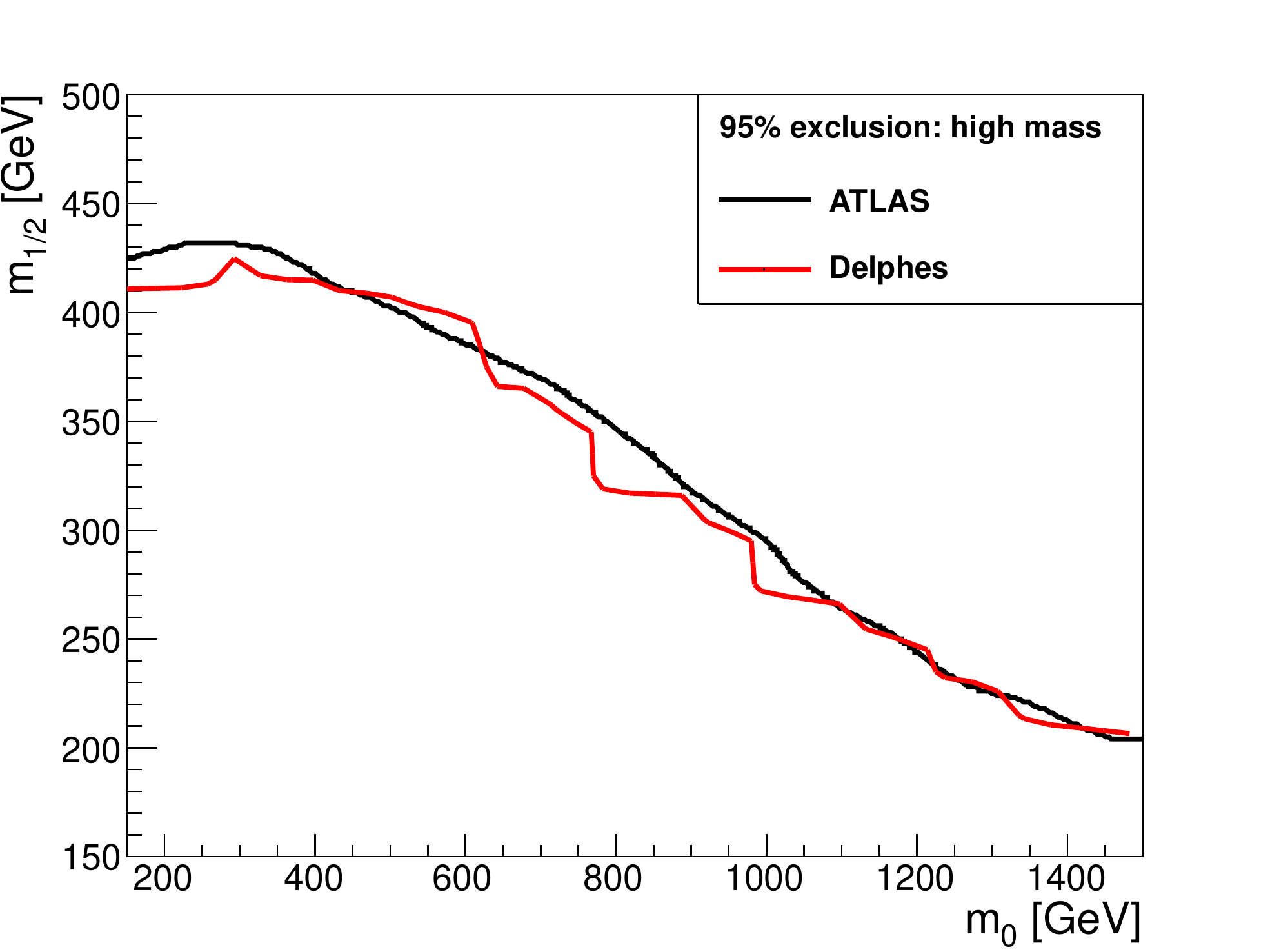}}
\caption{\label{0lep_valid} \footnotesize{Comparison of our 95\% C.L. exclusion contours with those of ATLAS in the $\tan\beta=10$, $A_0=0$, $\mu>0$ slice of the CMSSM for the 0-lepton search. 
}} 
\vspace{15pt}
\subfigure[7j55]{\includegraphics[width=0.31\textwidth]{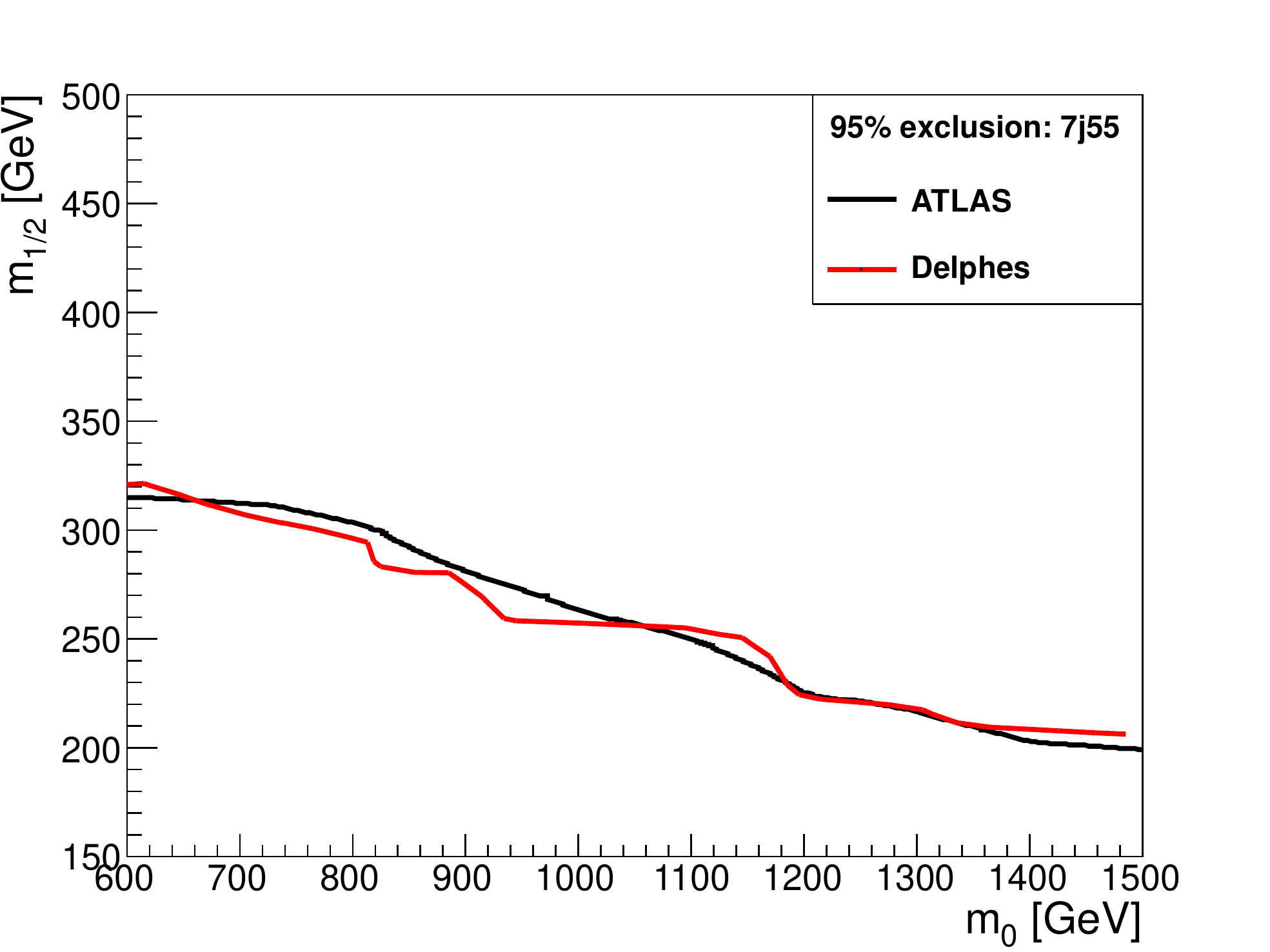}}
\hspace{-5pt}
\subfigure[8j55]{\includegraphics[width=0.31\textwidth]{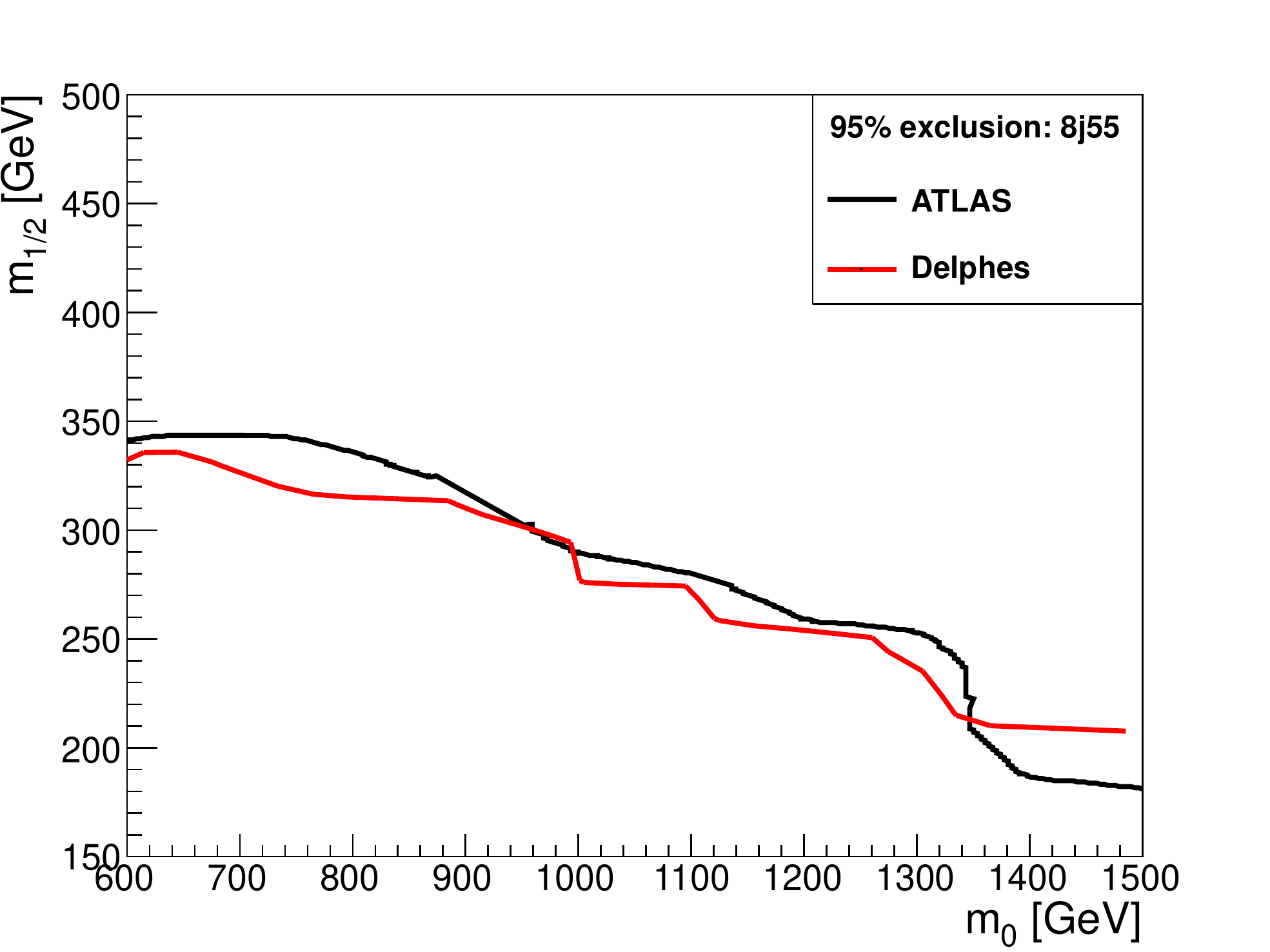}}
\hspace{-5pt}
\subfigure[6j80]{\includegraphics[width=0.31\textwidth]{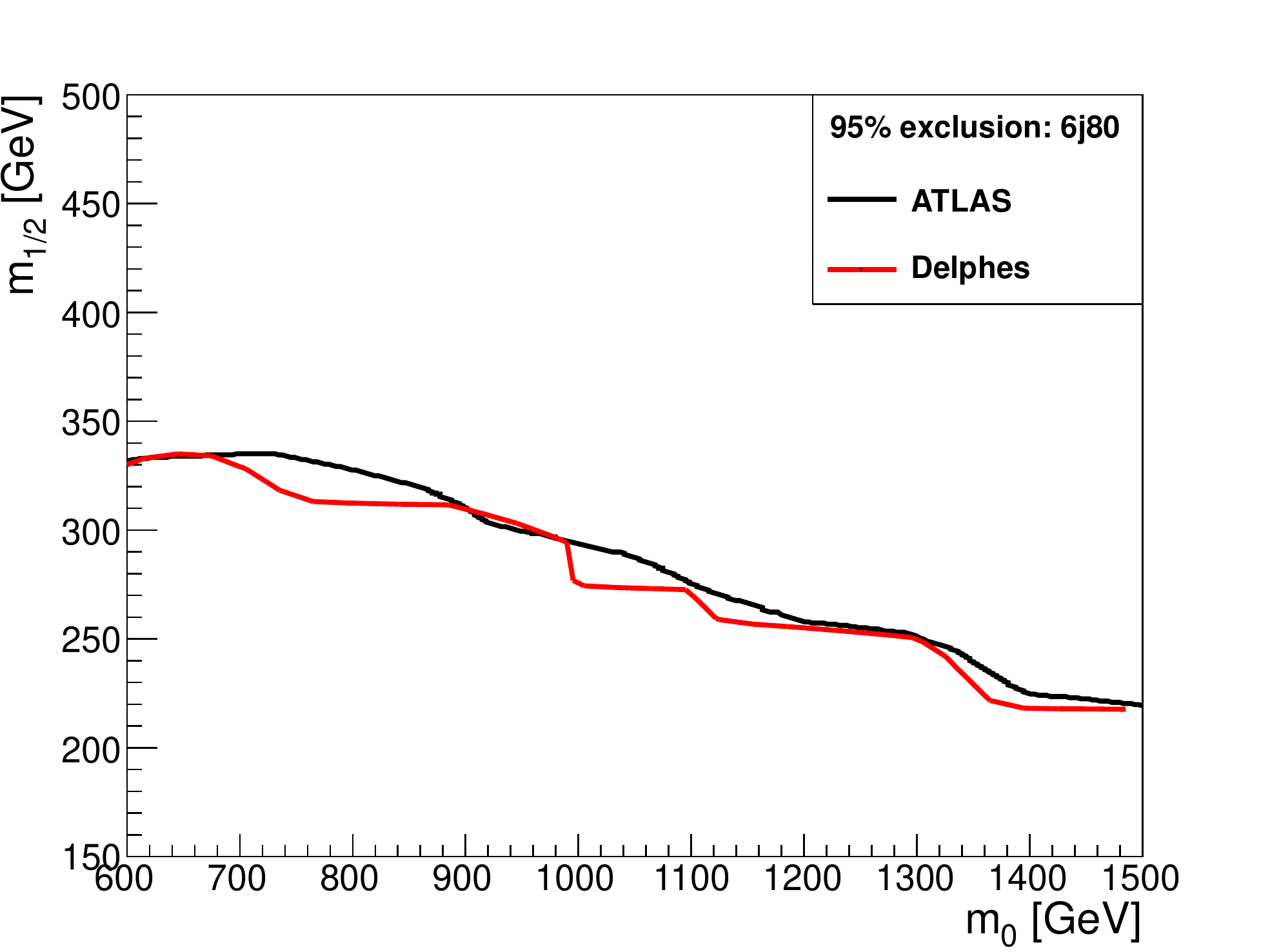}}
\\
\vspace{-10pt}
\subfigure[7j80]{\includegraphics[width=0.31\textwidth]{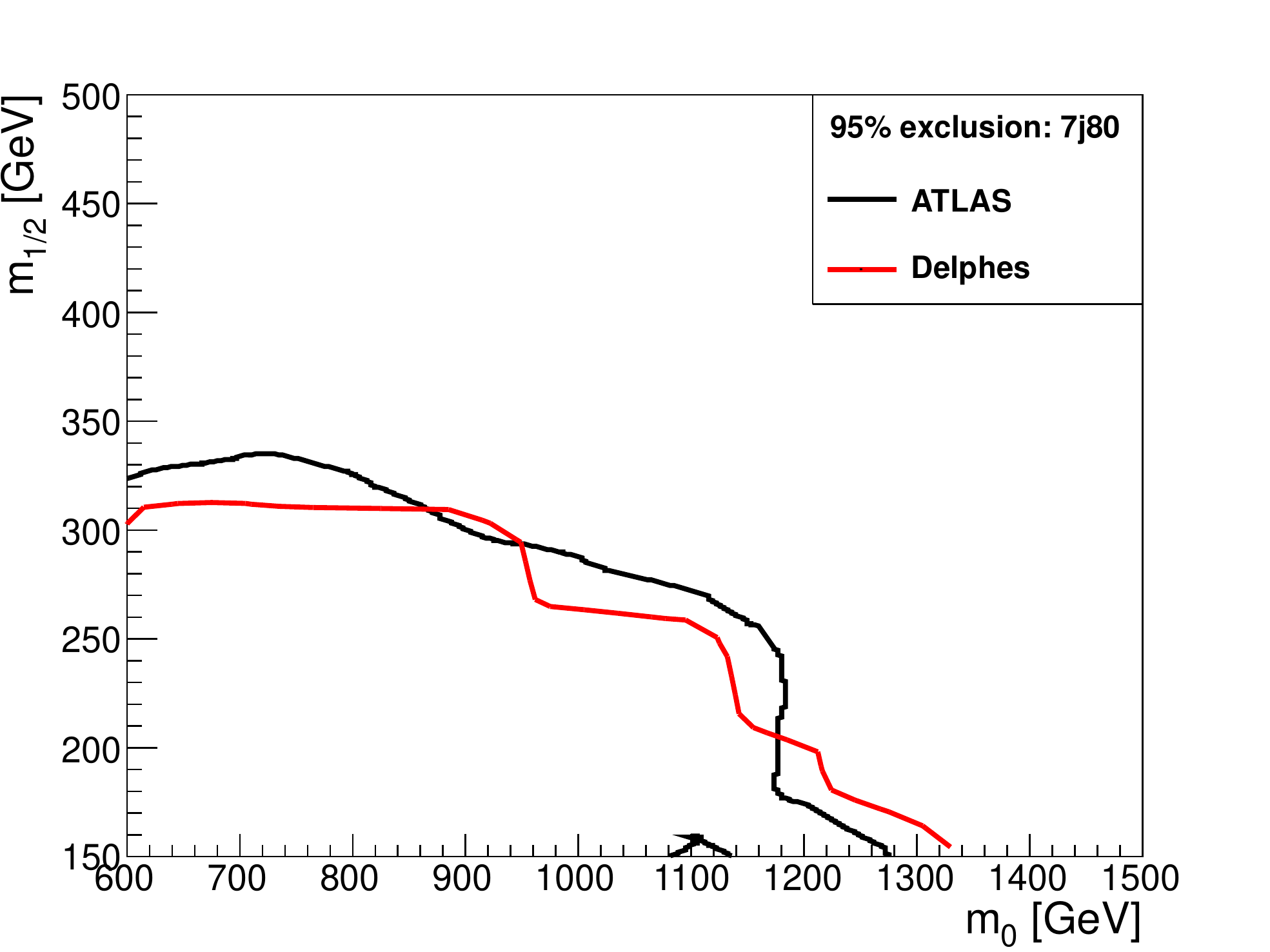}}
\caption{\label{multijets_valid} \footnotesize{Comparison of our 95\% C.L. exclusion contours with those of ATLAS in the $\tan\beta=10$, $A_0=0$, $\mu>0$ slice of the CMSSM for the multijets search. 
}}
\end{center}
\end{figure}

\appendix
\section{Validation of our simulation}

In this appendix we present the validation of our simulation of the ATLAS 0-lepton and multijets searches. For each of the signal region in those searches we determine the 95\% C.L. exclusion region in the $\tan\beta=10$, $A_0=0$, $\mu>0$ slice of the CMSSM and compare to the corresponding exclusion region obtained by ATLAS. Such comparisons are shown in Figures \ref{0lep_valid} and \ref{multijets_valid} for the 0-lepton search and the multijets search, respectively. It can be seen that, for each signal region, our exclusion contour is similar to the ATLAS one so we conclude that our approximation is reasonable.

%%%%%%%%%%%%%%%%%%%%%%%%%%%%%%%%%%%%%%%%%%%%%%%%%%%%%%%%%%
%%%%%%%%%%%%%%%%%%%%%%%%%%%%%%%%%%%%%%%%%%%%%%%%%%%%%%%%%%

\end{document}